\documentclass[aps,prx,a4paper,longbibliography,notitlepage,superscriptaddress,reprint,nofootinbib
]{revtex4-2}

\usepackage{amsmath,amssymb,amsfonts} 
\usepackage{mathrsfs} 
\usepackage{mathtools} 
\usepackage{color}
\usepackage[usenames,dvipsnames]{xcolor}
\usepackage{graphicx,float}
\usepackage[colorlinks, linkcolor=NavyBlue, citecolor=NavyBlue, urlcolor=NavyBlue, breaklinks=true]{hyperref} 
\usepackage{soul}
\usepackage{siunitx}
\usepackage{braket, bm, dsfont}

\makeatletter
\renewcommand\frontmatter@abstractwidth{\dimexpr\textwidth\relax}
\makeatother

\makeatletter
\newcommand{\fmarki}{*}
\newcommand{\fmarkii}{\ensuremath{\dagger}}
\newcommand{\fmarkiii}{\ensuremath{\ddagger}}
\def\@fnsymbol#1{{\ifcase#1\or \fmarki\or \fmarkii\or \fmarkiii \else\@ctrerr\fi}}
\makeatother

\renewcommand{\fmarki}{$\dagger$}
\renewcommand{\fmarkii}{$\ddagger$}
\renewcommand{\fmarkiii}{$\mathsection$}

\begin{document}

\title{
Entangling microwaves with optical light
}
\author{Rishabh Sahu$^\star$}
\email{rsahu@ist.ac.at}
\author{Liu Qiu$^\star$}
\email{liu.qiu@ist.ac.at}
\author{William Hease}
\author{Georg Arnold}
\affiliation{Institute of Science and Technology Austria, Am Campus 1, 3400 Klosterneuburg, Austria}
\author{Yuri Minoguchi}
\author{Peter Rabl}
\affiliation{Vienna Center for Quantum Science and Technology, Atominstitut, TU Wien, 1040 Vienna,
Austria}
\author{Johannes M. Fink}
\email{jfink@ist.ac.at}
\affiliation{Institute of Science and Technology Austria, Am Campus 1, 3400 Klosterneuburg, Austria}
\date{\today}

\begin{abstract}
\noindent 
Entanglement is a genuine quantum mechanical property and the key resource in currently developed quantum technologies. 
Sharing this fragile property between superconducting microwave circuits and optical or atomic systems would enable new functionalities but has been hindered by the tremendous energy mismatch of $\sim10^5$ and the resulting mutually imposed loss and noise.
In this work we create and verify entanglement between microwave and optical fields in a millikelvin environment. 
Using an optically pulsed superconducting electro-optical device, we deterministically prepare an itinerant microwave-optical state that is squeezed by $0.72^{+0.31}_{-0.25}$\,dB and violates the Duan-Simon separability criterion by $>5$ standard deviations. 
This establishes the long-sought non-classical correlations 
between superconducting circuits and telecom wavelength light with wide-ranging implications for hybrid quantum networks in the context of modularization, scaling, sensing and cross-platform verification.
\end{abstract}

\maketitle
\def\thefootnote{$\star$}
\footnotetext{These authors contributed equally to this work.}
\def\thefootnote{\arabic{footnote}}

Over the past decades we have witnessed spectacular progress in our capabilities to manipulate and measure genuine quantum mechanical properties, such as quantum superpositions and entanglement, in a variety of physical systems.
These techniques serve now as the basis for the development of quantum technologies, where the demonstration of quantum supremacy with tens of superconducting qubits~\cite{arute2019}, an ultra-coherent quantum memory with nuclear spins~\cite{zhong2015}, and distributed quantum entanglement over tens of kilometers using optical photons~\cite{yu2020} represent just a few of the highlights that have already been achieved. 
Going forward, combining these techniques~\cite{Xiang2013,kurizki2015,Clerk2020} will enable the realization of general-purpose quantum networks, where remote quantum nodes, capable of storing and processing quantum information, seamlessly communicate with each other by distributing entanglement over optical channels~\cite{duan2001}. 
Aspects of this approach have already been adopted to connect and entangle various quantum platforms remotely, involving single atoms, ions, atomic ensembles, quantum dots, rare-earth ions and nitrogen-vacancy centers~\cite{wei2022}.
However, such long-distance quantum connectivity is considerably more difficult to achieve with other promising platforms, such as semiconductor spin qubits or local cryogenic networks of superconducting circuits~\cite{magnard2020,zhong2021}, 
where no natural interface to room temperature noise-resilient optical photons is available. 

To overcome this limitation, a lot of effort is currently focused on the development of coherent quantum transducers between microwave and optical photons~\cite{han2021,mirhosseini2020,sahu2022,brubaker2022,kumar2022,weaver2022}. 
Direct noiseless conversion of a quantum state typically relies on a beam splitter process, 
where a strong driving field mediates the conversion between weak microwave and optical signals - a deterministic approach 
with exceptionally stringent requirements on conversion efficiency and added classical noise that are still out of reach.
Alternatively, the direct generation of quantum-correlated microwave-optical photon pairs can also be used as a resource for quantum teleportation and entanglement distribution in the continuous and discrete variable domain~\cite{takeda2013,Fedorov2021,hermans2022}.

\begin{figure}
	\centering
	\includegraphics[scale=1.2]{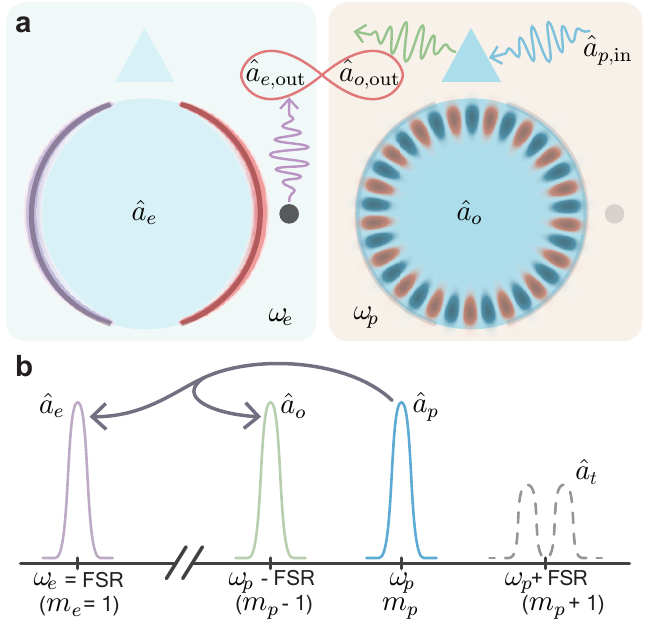}
	\caption{\textbf{Physical and conceptual mode configuration.}		 
\textbf{a}, Simulated microwave (left) and optical (right) mode distribution with azimuthal number $m_e=1$ and $m_o=17$ (for illustration, experimentally $m_o\approx$ \num{20000}). Phase matching is fulfilled due to the condition $m_o=m_p-m_e$ and entanglement is generated and verified between the out-propagating microwave field $\hat{a}_{e,\mathrm{out}}$ and the optical Stokes field $\hat{a}_{o,\mathrm{out}}$. 
\textbf{b}, Sketch of the density of states of the relevant modes. Under the condition $\omega_p-\omega_o=\omega_e$ the strong pump tone in $\hat{a}_p$ efficiently produces entangled pairs of microwave and optical photons in $\hat{a}_e$ and $\hat{a}_o$ via spontaneous parametric downconversion. 
Frequency up-conversion is suppressed via hybridization of the anti-Stokes mode $\hat{a}_t$ with an auxiliary mode.
	}
	\label{fig:fig1}
\end{figure}

In this paper we use an ultra-low noise cavity electro-optical device to generate such non-classical correlations in a deterministic protocol~\cite{rueda2019}. It consists of a 5 millimeter diameter, 150\,$\mu$m thick lithium niobate optical resonator placed inside a superconducting aluminum microwave cavity at a temperature of 7\,mK, described in detail in Ref.~\cite{hease2020}. As depicted in Fig.~\ref{fig:fig1}A, the microwave mode $\hat{a}_e$ is co-localized with and electro-optically coupled to the optical whispering gallery modes at $\omega_o/(2\pi)\approx 193.46$\,THz via the Pockels effect.
We match the tunable microwave resonance frequency $\omega_e/(2\pi)$ to the free spectral range (FSR) of \SI{8.799}{GHz} to realize a triply-resonant system~\cite{Ilchenko2003,Fan2018} with the interaction Hamiltonian,
\begin{equation}
	\hat{H}_\text{int} = \hbar g_0  \hat{a}_p \hat{a}_e^\dagger  \hat{a}_o^\dagger + \textrm{h.c.},
\end{equation}
with $g_0$ the vacuum electro-optical coupling rate, and $\hat{a}_p$ ($\hat{a}_o$) the annihilation operator of the optical pump (Stokes) mode~\cite{Tsang2010}.
Here we have ignored the interaction with the suppressed optical anti-Stokes mode $\hat{a}_t$, as shown in Fig.~\ref{fig:fig1}b (see supplementary information).

In this sideband suppressed situation, efficient two-mode squeezing is achieved with a strong resonant optical pump tone, yielding the simple effective Hamiltonian, $\hat{H}_\text{eff} = \hbar g_0 \sqrt{\bar{n}_p} ( \hat{a}_e^\dagger \hat{a}_o^\dagger + \hat{a}_e \hat{a}_o)$, where $\bar{n}_p = \braket{\hat{a}_p^\dagger \hat{a}_p}$ is the mean intra-cavity photon number of the optical pump mode. Deterministic continuous-variable (CV) entanglement between the out-propagating microwave and optical field can be generated below the parametric instability threshold ($\textit{C}<1$) in the quantum back-action dominated regime, where the quantum noise exceeds the microwave thermal noise. Here $C=4\bar{n}_p g_0^2/(\kappa_e\kappa_o)$ is the cooperativity with the vacuum coupling rate $g_0/2\pi\approx\,$\SI{37}{Hz}, and the total loss rates of the microwave and optical Stokes modes $\kappa_{e}/2\pi\approx\,$\SI{11}{MHz} and $\kappa_{o}/2\pi\approx\,$\SI{28}{MHz}. The required ultra-low noise operation is achieved despite the required high power optical pump due to slow heating of this millimeter sized device~\cite{hease2020}. 

In the following we characterize the microwave and optical output fields via the dimensionless quadrature pairs $X_j$ and $P_j$ ($j=e,o$ for microwave and optics), which satisfy the canonical commutation relations $\left[X_j, P_j\right]=i$.
A pair of Einstein-Podolsky-Rosen (EPR)-type operators $X_{+}=\frac{1}{\sqrt{2}}\left(X_e + X_o\right)$ and $P_{-}=\frac{1}{\sqrt{2}}\left(P_e - P_o\right)$ are then constructed, and the microwave and optical output fields are entangled, if the variance of the joint operators is reduced below the vacuum level, i.e. $\Delta_\text{EPR}^{-}=\left\langle X_{+}^2\right\rangle+\left\langle P_{-}^2\right\rangle<1$. This is commonly referred to as the Duan-Simon criterion \cite{duan2000, simon2000}, which we apply to each near-resonant frequency component of the two-mode squeezed output mode (see SI). 

\begin{figure*}
	\centering
	\includegraphics[scale=0.75]{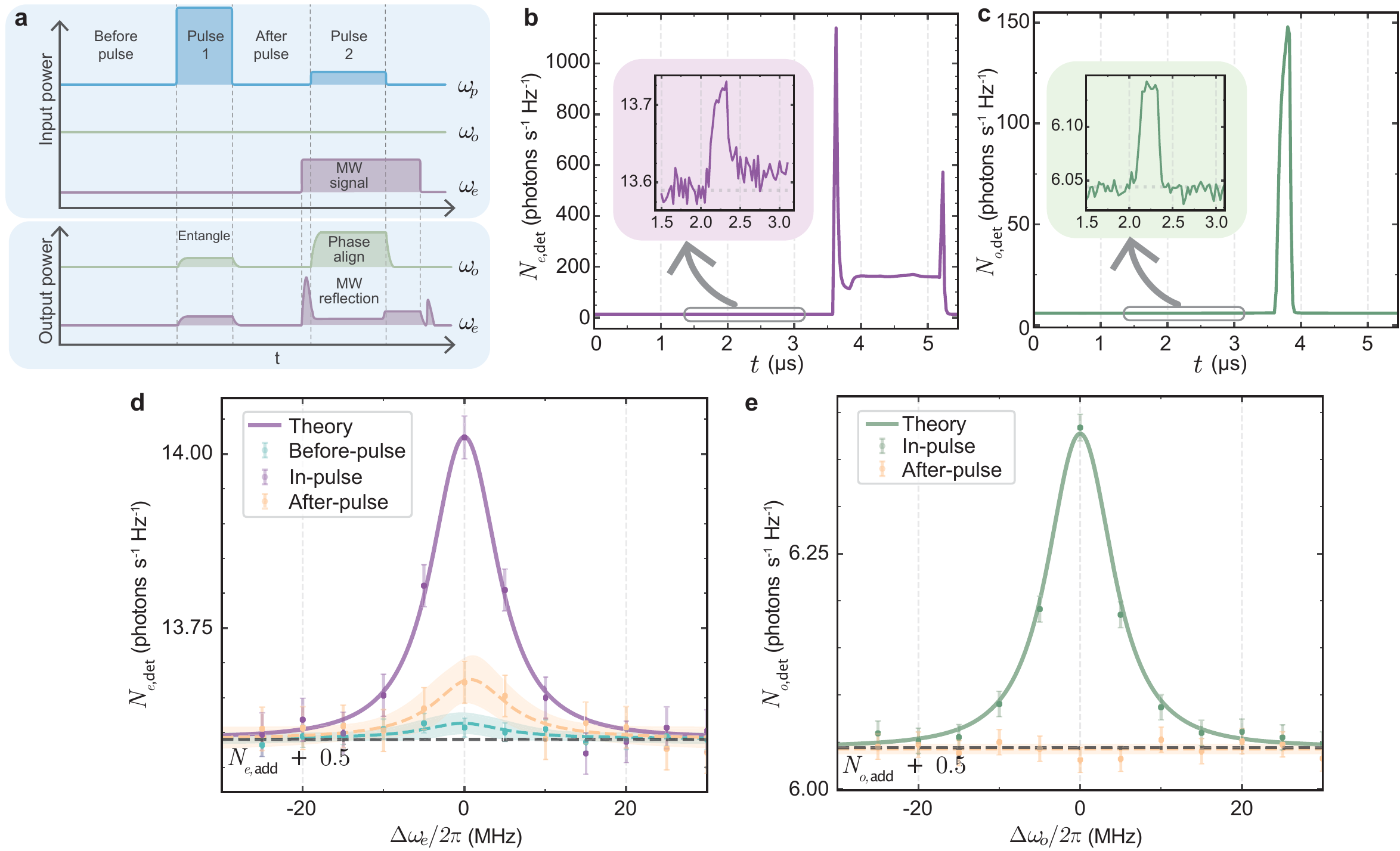}
	\caption{\textbf{Measurement sequence and noise powers.} 
\textbf{a}, Schematic pulse sequence of a single measurement. The optical pulse 1 is applied at $\omega_p$ and amplifies the vacuum (and any thermal noise) in the two modes $\hat{a}_e$ and $\hat{a}_o$, thus generating the SPDC signals. \SI{1}{\micro s} later, a second optical pump with about 10 times lower power is applied together with a coherent microwave pulse at $\omega_e$. The microwave photons stimulate the optical pump to down-convert, which generates a coherent pulse in the $\hat{a}_o$ mode that is used to extract slow LO phase drifts. 
\textbf{b} and \textbf{c}, Measured output power in the $\hat{a}_e$ and $\hat{a}_o$ mode in units of photons per second in a 1 Hz bandwidth and averaged over a million experiments. The SPDC signals are shown in the insets with the dashed gray lines indicating the calibrated detection noise floor $N_{j,\text{add}}+0.5$. 
\textbf{d}, Corresponding microwave output power spectral density vs. $\Delta\omega_e=\omega-\omega_e$ centered on resonance right before the entanglement pulse, during the pulse and right after the pulse, as indicated in panel \textbf{a}. 
Yellow and green dashed lines are fits to a Lorentzian function, which yields the microwave bath occupancies before and after the entangling pulse. Error bars represent the $2\sigma$ statistical standard error and the shaded regions represent the 95\% confidence interval of the fit.  
(\textbf{e}), Corresponding optical output power spectral density vs. $\Delta\omega_o=\omega_o-\omega$ during and after the entanglement pulse, both normalized to the measured noise floor before the pulse. The in-pulse noise spectra in panels \textbf{d} and \textbf{e} are fit jointly with theory, which yields $C=0.18\pm0.01$ and $\bar{n}_{e,{\mathrm{int}}}=0.07\pm0.03$.
}
	\label{fig:fig2}
\end{figure*}

For efficient entanglement generation, we use a \SI{250}{ns} long optical pump pulse ($\approx$ \SI{244}{mW}, 
$C \approx 0.18$, $\bar{n}_p\approx1.6\times10^{10}$) at a \SI{2}{Hz} repetition rate, cf. pulse 1 in Fig.~\ref{fig:fig2}a. 
The entangled output optical signal is filtered via a Fabry-Perot cavity to reject the strong pump. The entangled microwave output is amplified with a high-electron-mobility transistor amplifier. Both outputs are down-converted to an intermediate frequency of \SI{40}{MHz} with two local oscillators (LO) and the four quadratures are extracted from heterodyne detection. 
Long-term phase stability between the two LOs is achieved via extracting the relative phase drift by means of a second phase alignment pump pulse that is applied \SI{1}{\micro s} after each entanglement pulse, together with a coherent resonant microwave pulse, shown in Fig. \ref{fig:fig2}(a). This generates a high signal-to-noise coherent optical signal via stimulated parametric down-conversion and allows for aligning the phase of each individual measurement.

Figure~\ref{fig:fig2}b(c) shows the time-domain average power over one million averages for the on-resonant microwave (optics) signal with a spectrally under-sampled \SI{40}{MHz} bandwidth (hence not revealing the full pulse amplitude).
The two insets show the microwave (optical) signal from spontaneous parametric down-conversion (SPDC) due to pulse 1, cf.~Fig.~\ref{fig:fig2}a, with an emission bandwidth of $\approx$\,\SI{10}{MHz}. 
The larger signals during the second half of the experiment are the reflected microwave pulse and the generated optical tone (due to pulse 2) that is used for LO phase alignment. 
The raw power measurements are divided by the measurement bandwidth and rescaled such that the off-resonant response matches the noise photon number $N_{j,\mathrm{add}}+0.5$ of the measurement setup.
$N_{e,\mathrm{add}} = 13.1\pm0.4$ ($2\sigma$ errors throughout the paper) due to loss and amplifier noise and 
$N_{o,\mathrm{add}} = 5.5\pm0.2$ due to optical losses are carefully determined using noise thermometry of a temperature controlled \SI{50}{\ohm} load and 4-port calibration, respectively (see SI). Using this procedure ensures that the reported photon number units correspond to the signals at the device outputs.

We continue the analysis in the frequency domain by calculating the Fourier transform of each measurement for three separate time intervals - before (2\,$\mu$s), during (200\,ns) and right after the entangling pump pulse (500\,ns).
Figure~\ref{fig:fig2}d shows the resulting average microwave noise spectra for all three time intervals with corresponding fit curves (dashed lines) and theory (solid line). Before and after the pump pulse, the on-resonant microwave output field takes on values above the vacuum level, with fitted intrinsic microwave bath occupancies $\bar{n}_{e,{\mathrm{int}}}=0.03\pm0.01$ and $0.09\pm0.03$, respectively.
By fitting additional power dependent measurements, we independently verify that the observed noise floor corresponds to a waveguide bath occupancy of only $\bar{n}_{e,{\mathrm{wg}}} = 0.001\pm0.002$ at the very low average pump power of $\approx$\,\SI{0.12}{\micro W} used in this experiment (see SI).
The measured noise floor therefore corresponds to the shot noise equivalent level $N_{e,\text{add}}+0.5$ (gray dashed lines).
Similarly, Fig.~\ref{fig:fig2}e shows the obtained average optical noise spectra during and after the pump, referenced to the measured shot noise level before the pulse. 
As expected, there is no visible increase of the optical noise level after the pulse.

During the pump pulse, an approximately Lorentzian shaped microwave and optical power spectrum are generated via the SPDC process. We perform a joint fit of the microwave and optical power spectral density during the pulse using a 5-mode theoretical model that includes the effects of measurement bandwidth. In this model, the in-pulse microwave bath occupancy 
$\bar{n}_{e,{\mathrm{int}}}=0.07\pm 0.03$
and the cooperativity $C = 0.18\pm 0.01$ are the only free fit parameters.
Here the narrowed microwave linewidth $\kappa_{e,{\mathrm{eff}}}/2\pi = 9.8\pm1.8\, \mathrm{MHz}$ (taken from a Lorentzian fit) agrees with coherent electro-optical dynamical back-action~\cite{qiu2022}, where $\kappa_{e,{\mathrm{eff }} } = (1-C) \kappa_{e}$.
We conclude that this cavity electro-optical device is deep in the quantum back-action dominated regime, a prerequisite for efficient microwave-optics entanglement generation.

\begin{figure*}
	\centering
	\includegraphics[scale=0.5]{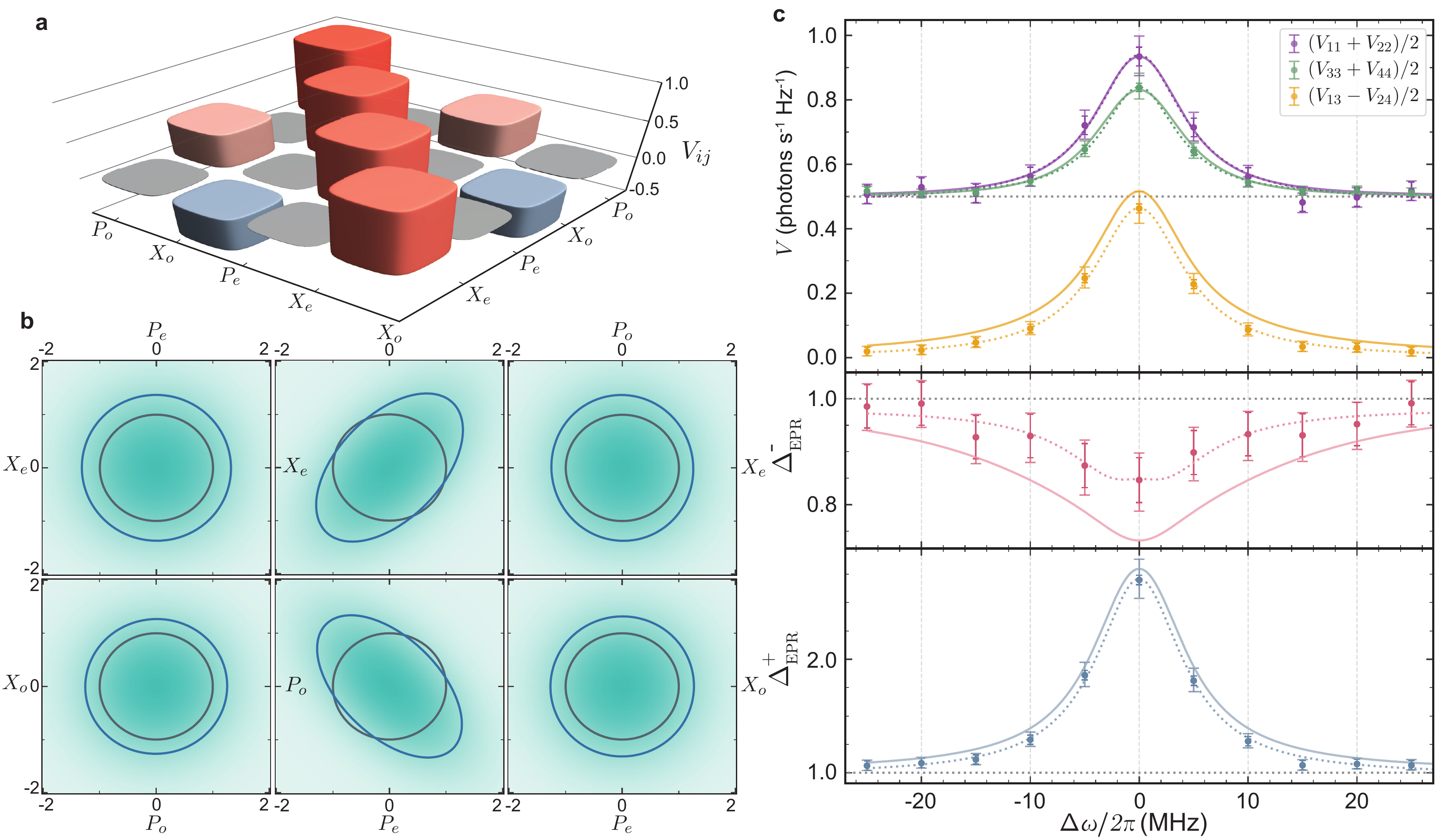}
	\caption{\textbf{Characterization of the two-mode squeezed state.} 
\textbf{a}, Measured covariance matrix $V_{ij}$ in its standard form plotted for $\Delta\omega_j=0$ based on 925000 measurements.
\textbf{b}, Corresponding Wigner function marginals of different output quadrature pairs in comparison to vacuum. The contours in blue (grey) represent the $1/e$ fall-off from the maximum for the measured state (vacuum). Middle two plots show two-mode squeezing below the vacuum level in the diagonal and off-diagonal directions. 
\textbf{c}, Top panel, the measured average microwave output noise $\bar{V}_{11} = (V_{11}+V_{22})/2$ (purple), the average optical output noise $\bar{V}_{33} = (V_{33}+V_{44})/2$ (green) and the average correlations $\bar{V}_{13} = (V_{11}-V_{24})/2$ (yellow) as a function of the measurement detunings.
The solid lines represent the joint theory fit 
and the dashed lines are individual Lorentzian fits to serve as a guide to eye. 
The middle (bottom) panel shows
two-mode squeezing in red (anti-squeezing in blue) calculated from the top panels as $\Delta_\text{EPR}^\pm =\bar{V}_{11}+\bar{V}_{33}\pm 2\bar{V}_{13}$. 
The darker color error bars represent the $2\sigma$ statistical error and the outer (faint) $2\sigma$ error bars also include the systematic error in calibrating the added noise of the measurement setup. 
		}
	\label{fig:fig3}
\end{figure*}

For each frequency component the bipartite Gaussian state of the propagating output fields can be fully characterized by the 
 $4\times4$ covariance matrix (CM) $V_{ij} = \braket{\delta u_i \delta u_j + \delta u_j \delta u_i}/2$, where $\delta u_i = u_i - \braket{u_i}$ and $u \in \{X_e, P_e, X_o, P_o\}$ (see SI).
The diagonal elements in $V$ correspond to the individual output field quadrature variances in dimensionless units. These are obtained from the measured variances after subtracting the measured detection noise offsets shown in Fig.~\ref{fig:fig2}, i.e. 
$V_{ii}(\Delta\omega) 
= V_{ii,\mathrm{meas}}(\Delta\omega_i) - N_{i,{\textrm{add}}}$.
The obtained CM from the data in Fig.~\ref{fig:fig2}
at $\Delta\omega=0$ is shown in Fig.~\ref{fig:fig3}a in its standard form. It corresponds to the quantum state 
of the propagating modes in the coaxial line and the coupling prism attached to the device output, i.e. 
before setup losses or amplification incur.  
The non-zero off-diagonal elements indicate strong correlations between microwave and optical quadratures.

The two-mode squeezed quadratures are more intuitively visualized in terms of the quasi-probability Wigner function,
\begin{equation}
	W(\bold{u}) = \frac{\exp[-\frac{1}{2}\bold{u}V^{-1}\bold{u}^T]}{\pi^2\sqrt{\text{det(V)}}},
\end{equation}
where $\bold{u} = (X_e, P_e, X_o, P_o)$. 
Different marginals of this Wigner function are shown in Fig.~\ref{fig:fig3}b, 
where the ($X_e$,$X_o$) and ($P_e$,$P_o$) marginals show two-mode squeezing in the diagonal and off-diagonal directions.
The two cross-quadrature marginals show a slightly different amount of squeezing, which is due to the statistical uncertainty in the measured CM. 

Figure~\ref{fig:fig3}c shows the amount of two-mode squeezing between microwave and optical photon pairs. 
Correlations are observed at $\Delta\omega_j=\pm(\omega-\omega_j)$ around the resonances
due to energy conservation in the SPDC process (see SI).  
The averaged microwave quadrature variance (purple dots) $\bar{V}_{11} = (V_{11}+V_{22})/2$ and the averaged optics quadrature variance (green dots) $\bar{V}_{33} = (V_{33}+V_{44})/2$ are shown in the
top panel along with the prediction from our five-mode theory (solid line) and a simple fit to a Lorentzian function (dashed line), showing perfect agreement.
Measured microwave-optical correlations (yellow dots) $\bar{V}_{13} = (V_{13}-V_{24})/2$ and the Lorentzian fit (dashed line) lie slightly below the theoretical prediction (solid line), which we assign to remaining imperfections in the phase stability (see SI). 

The bottom two panels of Fig.~\ref{fig:fig3}c show the squeezed and anti-squeezed joint quadrature variances $\Delta_\text{EPR}^{\mp}=\bar{V}_{11}+\bar{V}_{33}\mp2\bar{V}_{13}$ (red and blue color respectively). 
We observe two-mode squeezing below the vacuum level, i.e. $\Delta_\text{EPR}^{-}<1$, with a bandwidth close to the effective microwave linewidth. 
The maximal on-resonant two-mode squeezing is $\Delta_\text{EPR}^{-} = 0.85^{+0.05}_{-0.06}$ (2$\sigma$, 95\% confidence) for $\sim$1 million pulses with $\bar{V}_{11}=0.93$, $\bar{V}_{33}=0.84$ and $\bar{V}_{13}=0.46$. Hence, this deterministically generated microwave-optical state violates the Duan-Simon separability criterion by 
$>5\sigma$. Note that this error also takes into account systematics in the added noise calibration used for scaling the raw data (see SI).
These values correspond to a state purity of $\rho = 1/(4\sqrt{\text{det}[V]}) = 0.44$ 
and demonstrate microwave-optical entanglement between output photons with a logarithmic negativity of $E_N = \SI{0.17}{}$. The supplementary material contains substantial additional data for longer pulses and varying optical pump power, which corroborates the presented results and findings, albeit with lower statistical significance for each individual pump configuration (see SI).

\subsection*{Conclusions and outlook}
In conclusion, we have demonstrated deterministic quantum entanglement between propagating microwave and optical photons,thus establishing a non-classical communication channel between circuit quantum electrodynamics and quantum photonics. 
Our device can readily be used for probabilistic heralding assisted protocols \cite{duan2001,zhong2020,Krastanov2021} to mitigate optical setup losses and extend the entanglement to room temperature fiber optics. 
We expect that the pulse repetition rate can be increased by orders of magnitude with improved thermalization, higher microwave and optical quality factors, and electro-optic coupling enhancements that reduce the required pump power and the associated thermal load. 
Coupling efficiency improvements will allow for higher levels of two-mode squeezing and facilitate also deterministic entanglement distribution schemes \cite{agusti2022}, teleportation-based state transfer
~\cite{rueda2019,wu2021} and quantum-enhanced remote detection~\cite{barzanjeh2015}. 
Being fully compatible with superconducting qubits in a millikelvin environment
such a device will facilitate the integration of remote superconducting quantum processors into a single coherent optical quantum network. This is not only relevant for modularization and scaling~\cite{Bravyi2022,Ang2022}, but also for efficient cross-platform verification of classically intractable quantum processor results~\cite{Knorzer2022}. 

\section*{Acknowledgments}
L.Q. acknowledges fruitful discussions with Jie Li and David Vitali.
This work was supported by the European Research Council under grant agreement no. 758053 (ERC StG QUNNECT) and the European Union’s Horizon 2020 research and innovation program under grant agreement no. 899354 (FETopen SuperQuLAN). L.Q. acknowledges generous support from the ISTFELLOW programme. W.H. is the recipient of an ISTplus postdoctoral fellowship with funding from the European Union’s Horizon 2020 research and innovation program under the Marie Sk\l{}odowska-Curie grant agreement no. 754411. G.A. is the recipient of a DOC fellowship of the Austrian Academy of Sciences at IST Austria. J.M.F. acknowledges support from the Austrian Science Fund (FWF) through BeyondC (F7105) and the European Union's Horizon 2020 research and innovation programs under grant agreement No 862644 (FETopen QUARTET). \\

\section*{Author contributions}
RS, WH, LQ, and GA worked on the setup.
RS and LQ performed measurements.
LQ and RS did the data analysis.
LQ developed the theory with contributions from RS, YM and PR. 
RS and LQ wrote the manuscript with contributions from all authors. 
JMF supervised the project. \\

\section*{Data availability statement}
All data and code used to produce the figures in this manuscript will be made available on Zenodo. \\

\bibliography{Final_jf2}

\bibliographystyle{apsrev4-2}

\end{document}


\title{Supplementary Information for: "Entangling microwaves with optical light"}
\author{Rishabh Sahu}
\thanks{These two authors contributed equally}
\author{Liu Qiu}
\thanks{These two authors contributed equally}
\author{William Hease}
\author{Georg Arnold}
\affiliation{Institute of Science and Technology Austria, am Campus 1, 3400 Klosterneuburg, Austria}%
\author{Yuri Minoguchi}
\author{Peter Rabl}
\affiliation{Vienna Center for Quantum Science and Technology, Atominstitut, TU Wien, 1040 Vienna,
Austria}
\author{Johannes M. Fink}
\affiliation{Institute of Science and Technology Austria, am Campus 1, 3400 Klosterneuburg, Austria}%
\date{\today}

\maketitle

\onecolumngrid
\tableofcontents
\addtocontents{toc}{~\hfill\textbf{Page}\par}
\newpage

\begin{table}[h!]
	\begin{tabular}{|c|p{14cm}|}
		\hline
		\multicolumn{2}{|c|}{\textbf{Introduced in Main Text}} \\ \hline
		$\hat{a}_e$                                                                              & microwave mode (annihilation  operator)                                                                                \\ \hline
		$m_e$                                                                              & microwave mode azimuthal number, $m_e=1$                                                                                  \\ \hline
		$\omega_e$                                                                          & microwave cavity frequency                                                               \\ \hline
		$\kappa_e$, $\kappa_{e,\mathrm{eff}}$, $\kappa_{e,\mathrm{in}}$, $\kappa_{e,\mathrm{0}}$& microwave total loss,  effective total loss, waveguide coupling and intrinsic loss rates \\ \hline
		$\bar{n}_{e,\mathrm{int}}, \bar{n}_{e,\mathrm{wg}}$                                 & microwave intrinsic and waveguide bath occupancy                                         \\ \hline
		$N_{e,\mathrm{add}}$                                                                &  added noise in the microwave detection                                                                    \\ \hline
		$\hat{a}_o$, $\hat{a}_p$, $\hat{a}_t$, $\hat{a}_{\mathrm{tm}}$                                                                               & optical Stokes, pump, anti-Stokes, and transverse-magnetic mode (annihilation  operator)                                                                      \\ \hline
		$\hat{a}_{e/o,\rm out}$   & microwave and optical output field from the device \\ \hline
		$m_p$         &
        optical pump mode azimuthal number, $m_p\sim20000$                             \\ \hline
		$\kappa_o$                                                                          & optical total loss rate                                                                  \\ \hline
		$N_{o,\mathrm{add}}$                                                                &  added noise in the optical detection                                                                      \\ \hline
		$\bar{n}_p$                                                                         & mean photon number of the optical pump mode                                              \\ \hline
		$g_0$                                                                               & electro-optical vacuum coupling rate                                                     \\ \hline
		$g$                                                                                   & photon enhanced electro-optical coupling rate ($g=\sqrt{ \bar{n}_p} g_0$\})                 \\ \hline
		$C$                                                                                   & cooperativity ( $ C= 4 g^2/\kappa_e\kappa_o$ )                                             \\ \hline
		$X_e$,$P_e$                                                                         & quadratures of the microwave output field                                                                    \\ \hline
		$X_o$,$P_o$                                                                         & quadratures of the optical Stokes output field                                                                     \\ \hline
		$V$                                                                         & covariance matrix of the bipartite Gaussian state, $V_{ij} = \braket{\Delta u_i \Delta u_j + \Delta u_j \Delta u_i}/2$, where $\Delta u_i = u_i - \braket{u_i}$ and $u \in \{X_e, P_e, X_o, P_o\}$.                                                               \\ \hline
		$N_{ii,{\textrm{add}}}$ & added noise in the quadrature variances measurements, $N_{11,{\textrm{add}}} = N_{22,{\textrm{add}}}=N_{e,{\textrm{add}}}$, $N_{33,{\textrm{add}}} = N_{44,{\textrm{add}}}=N_{o,{\textrm{add}}}$ \\ \hline
		$V_{ii,\mathrm{meas}}$ & diagonal covariance matrix elements from the calibrated measurement record, $V_{ii} = V_{ii,\mathrm{meas}} - N_{ii,{\textrm{add}}}$ \\ \hline
		$V_{11}$,$V_{22}$, $\bar{V}_{11}$                                                                         & quadrature variances of the microwave output field, $\bar{V}_{11} = \frac{V_{11}+V_{22}}{2}$                                                                \\ \hline
		$V_{33}$,$V_{44}$, $\bar{V}_{33}$                                                                         & quadrature variances of the optical Stokes output field, $\bar{V}_{33} = \frac{V_{33}+V_{44}}{2} $\\ \hline
		$V_{13}$,$V_{24}$, $\bar{V}_{13}$                                                                         & cross-correlation between microwave and optical quadratures, $\bar{V}_{13} = \frac{V_{13} -V_{24}}{2} $\\ \hline
		$\Delta^{\mp}_{\mathrm{EPR}}$                                                                         & squeezed and anti-squeezed joint quadrature variance between microwave and optical output field, $\Delta^{\mp}_{\mathrm{EPR}} =\bar{V}_{11} +  \bar{V}_{33} \mp \bar{V}_{13}$\\ \hline
		\multicolumn{2}{|c|}{\textbf{Introduced in Supplementary Information}} \\ \hline
        $J$                                                                         & coupling rate between the optical anti-Stokes mode and TM mode
        \\ \hline
		$\hat{a}_{e/o,\mathrm{in}}$                                                                         & input field (noise) operator for the microwave and optical mode\\ \hline		$\hat{a}_{e/o,\mathrm{0}}$                                                                         & noise operator for the microwave and optical intrinsic loss
        \\ \hline
		$\eta_{j}$                                                                         & external cavity coupling efficiency of individual mode, $j\in(e,o,p,t)$
        \\ \hline
        $\mathcal{G}(\omega)$              & spectral filter of the output field
		\\ \hline
		$\hat{A}(\Omega)$                                                                         & Fourier transform of operator $\hat{A}(t)$, $\hat{A}(\Omega) = \int\mathrm{d} t\, e^{i\Omega t}\hat{A}(t)$, $\hat{A}^{\dagger}(\Omega)=\int d t \hat{A}^{\dagger}(t) e^{i \Omega t}=[\hat{A}(-\Omega)]^{\dagger}$
		\\ \hline
        $\hat{X}(\omega_n)$ & $X$ quadrature of the output spectral mode, $\hat{X}(\omega_n) = \frac{1}{\sqrt{2}}\int_{-\infty}^\infty\mathrm{d}\omega\,
        \mathcal{G}(\omega_n - \omega)\hat{a}_{{\rm out}}(\omega) + \mathrm{h.c.}$  
		\\ \hline
        $\hat{P}(\omega_n)$ & $P$ quadrature of the output spectral mode, $\hat{P}(\omega_n) = \frac{1}{\sqrt{2}i}\int_{-\infty}^\infty\mathrm{d}\omega\,
        \mathcal{G}(\omega_n - \omega)\hat{a}_{{\rm out}}(\omega) + \mathrm{h.c.}$  
        \\ \hline
        $S_{\hat{A}\hat{B}}(\Omega)$                                                                         & Two-time correlation of two operators, $S_{\hat{A}\hat{B}}(\Omega) = \frac{1}{\sqrt{2\pi}}\int^{\infty}_{-\infty}\left<\hat{A}(t)\hat{B}(t')\right> e^{i\Omega t} \mathrm{d} t$ 
		\\ \hline
        $\Delta_{\rm LO}$      &
        local oscillator and signal frequency difference in heterodyne measurement, $\Delta_{\rm LO} = \omega_{\rm LO} - \omega_{\rm sig}$
        \\ \hline
        $I_{\rm out}(t), I_{\rm out}(\omega)$      &
        unitless output field in the equivelant heterodyne detection
        \\ \hline
        $S_{II}(\omega)$      &
		double-sided noise spectrum of the output field in the equivelant heterodyne detection
		\\ \hline
		$G_{\rm det}(\omega)$      &
		frequency dependent detection gain 
        in the heterodyne detection
		\\ \hline
        $\hat{I}_{X/P,{\rm det}}(\omega_n)$     &
		detected output photocurrent quadratures in  heterodyne detection, including detection gain
		\\ \hline
		$\hat{I}_{X/P,{\rm out}}(\omega_n)$      &
		unitless output field quadratures from $I_{\rm out}$ including added noise
		\\ \hline
		$\mathcal{D}(\omega)$      &
		covariance matrix of the detected quadratures \textit{from the heterodyne measurement record}
		\\ \hline
		$V_{\rm meas}(\omega)$      &
		covariance matrix of the total measured output field quadratures including added noise
		\\ \hline
		$\hat{X}_{+}(\omega)$      &
		joint quadrature of $\hat{X}_e(\omega)$ and $\hat{X}_o(-\omega)$, $\hat{X}_{+}(\omega) =(\hat{X}_e(\omega) + \hat{X}_o(-\omega))/\sqrt{2}$
		\\ \hline
		$\hat{P}_{-}(\omega)$      &
		joint quadrature of $\hat{P}_e(\omega)$ and $\hat{P}_o(-\omega)$, $\hat{P}_{-}(\omega) =(\hat{P}_e(\omega) - \hat{P}_o(-\omega))/\sqrt{2}$
		\\ \hline
		$E_N$      &
		logarithm negativity
		\\ \hline
		$\rho$      &
		state purity
        \\ \hline
		\end{tabular}
\end{table}

\newpage
\section{Theory}

\subsection{Covariance Matrix from Input-Output Theory}
\subsubsection{Quantum Langevin Equations}
Our cavity electro-optical (CEO) device consists of a millimeter-sized lithium niobate optical resonator in a 3-D superconducting microwave cavity at mK temperature~\cite{hease2020}.
The Pockels effect in lithium niobate allows for direct coupling between the microwave and optical whispering gallery modes with maximal field overlap.
The optical free spectral range (FSR) matches the microwave cavity frequency, with microwave azimuthal mode number $m_e = 1$.
As shown in Fig.~1 in the main text, resonant three-wave mixing between the microwave mode ($\hat{a}_e$) and three adjacent transverse-electric  (TE) optical modes, i.e. Stokes ($\hat{a}_o$), pump ($\hat{a}_p$), and anti-Stokes ($\hat{a}_t$) mode, arises via the cavity enhanced electro-optical interaction~\cite{Tsang2010,Tsang2011}.
In addition, the anti-Stokes mode is coupled to a transverse-magnetic (TM) optical mode ($\hat{a}_{\mathrm{tm}}$) of orthogonal polarization and similar frequency at rate of $J$~\cite{Rueda2016}.
This results in a total interaction Hamiltonian, 
\begin{equation}
	\hat{H}_I/\hbar = g_{0} (\hat{a}^\dagger_p \hat{a}_e \hat{a}_o  
	+  \hat{a}^\dagger_p \hat{a}^\dagger_e \hat{a}_{t} ) + J \hat{a}_t  \hat{a}_{\mathrm{tm}}^\dagger 
	  + h.c. ,
	\label{eq:H_I}
\end{equation}
with $g_0$ the vacuum electro-optical coupling rate.

For efficient entanglement generation, we drive the pump mode strongly with a short coherent input pulse $\bar{a}_{p,\mathrm{in}}(t)$ at frequency $\omega_p$~\cite{hease2020}, which results in a time-dependent mean intra-cavity field of the pump mode $\bar{a}_p (t)$,
\begin{equation}
\dot{\bar{a}}_p  = \left(i \Delta_p-\frac{\kappa_p}{2}\right)\bar{a}_p
+\sqrt{\eta_p \kappa_p } \bar{a}_{p, \mathrm{in}},
\end{equation}
where the pump tone is detuned from the pump mode by $\Delta_p = \omega_p - \omega_{o,p}$, with $\kappa_p$ and $\eta_p$ as the pump mode loss rate and external coupling efficiency.
\textit{In our experiments, we actively lock the laser frequency to the pump mode resonance, with $\Delta_p=0$.}

The presence of the strong pump field results in 
an effective interaction Hamiltonian,
\begin{equation}
	\hat{H}_{I,\mathrm{eff}}/\hbar = g ( \hat{a}_e\hat{a}_o  	+  \hat{a}_e\hat{a}^\dagger_{t} )  + J \hat{a}_t  \hat{a}_{\mathrm{tm}}^\dagger
		  + h.c.,
	\label{eq:H_eo}
\end{equation}
with multiphoton coupling rate $g = \bar{a}_p g_0$.
This includes the two-mode-squeezing (TMS) interaction between the Stokes and microwave mode, and beam-splitter (BS) interaction between the anti-Stokes mode and microwave mode, resulting in scattered Stokes and anti-Stokes sidebands that are located on the lower and upper side of the pump tone by $\Omega_e$ away.
Microwave-optics entanglement between the microwave and optical Stokes output field can be achieved via spontaneous parametric down-conversion (SPDC) process due to TMS interaction~\cite{qiu2022}, which is further facilitated by the suppressed anti-Stokes scattering due to the strong coupling between anti-Stokes and TM modes.
We can obtain the full dynamics of the intracavity fluctuation field \textit{in the rotating frame of the scattered sidebands and microwave resonance}, which can be described by the quantum Langevin equations (QLE),
\begin{eqnarray}
		\dot{\hat{a}}_e&=&-\frac{\kappa_e}{2} \hat{a}_e-i g\hat{a}_o^{\dagger} -i g^* \hat{a}_t
		+\sqrt{\eta_e \kappa_e} \delta \hat{a}_{e, \mathrm{in}} 
		+ \sqrt{\left(1-\eta_e\right) \kappa_e} \delta\hat{a}_{e, 0}, \\
		\dot{\hat{a}}_o&=&\left(i \delta_o-\frac{\kappa_o}{2}\right) \hat{a}_o-i g \hat{a}_e^{\dagger} 
		+\sqrt{\eta_o \kappa_o} \delta\hat{a}_{o, \mathrm{in}} + \sqrt{\left(1- \eta_o\right) \kappa_o} \delta\hat{a}_{o, \mathrm{0}}, \\
		\dot{\hat{a}}_t&=&\left(i \delta_t-\frac{\kappa_t}{2}\right) \hat{a}_t-i g^* \hat{a}_e - i J \hat{a}_{\mathrm{tm}} + \sqrt{\kappa_t} \delta\hat{a}_{t, \mathrm{vac}}, \\
		\dot{\hat{a}}_{\mathrm{tm}}&=&\left(i \delta_{\mathrm{tm}}-\frac{\kappa_{\mathrm{tm}}}{2}\right) \hat{a}_{\mathrm{tm}} - i J \hat{a}_t + \sqrt{\kappa_{\mathrm{tm}}} \delta\hat{a}_{{\mathrm{tm}}, \mathrm{vac}},
\end{eqnarray}
with $\kappa_{j}$ the total loss rate of the individual mode where $j \in (e, o, t,\mathrm{tm})$, and $\eta_{k}$ the external coupling efficiency of the input field where $k \in (e, o)$.
We note that, the optical light is only coupled to the TE modes via efficient prism coupling, with effective mode overlap $\Lambda$ factor included in $\eta_{o}$ for simplicity~\cite{hease2020}.  
$\delta_j$ corresponds to the frequency difference between mode $j$ and scattered sidebands, with $\delta_o = \omega_{o,p} - \omega_e  - \omega_o$ and $\delta_{t/\mathrm{tm}} = \omega_{o,p} + \omega_e  - \omega_{t/\mathrm{tm}}$, which are mostly given by FSR and $\omega_e$ mismatch, with additional contributions from optical mode dispersion and residual optical mode coupling. 
We note that, for resonant pumping, we have $\delta_o = - \delta_t$ in the case of absent optical mode dispersion and residual mode coupling.
In our experiments, we tune the microwave frequency to match the optical FSR, i.e. $\omega_e =  \omega_{o,p} - \omega_o$.

The equation of motion of all relevant modes may be represented more economically in the form
\begin{equation}
	\label{eq:QLE_vec}
	\dot{\boldsymbol{v}}(t) = M(t) \boldsymbol{v}(t) + K \boldsymbol{f}_{\mathrm{in}}(t),
\end{equation}
where we define the vectors of mode and noise operators
\begin{equation}
	\begin{aligned}
		\boldsymbol{v} &= (
		\hat{a}_e , \hat{a}_e^\dagger ,  \hat{a}_o, \hat{a}_o^\dagger ,  \hat{a}_t ,  \hat{a}_t^\dagger,  \hat{a}_{\mathrm{tm}},  \hat{a}_{\mathrm{tm}}^\dagger)^\top,\\
		\boldsymbol{f}_{\rm in}
		&=
		(
  \delta\hat{a}_{e,\mathrm{0}}, \delta\hat{a}_{e,\mathrm{0}}^\dagger,
  \delta\hat{a}_{e,\rm in}, \delta\hat{a}_{e,\rm in}^\dagger , 
  \delta\hat{a}_{o,\mathrm{0}} , \delta\hat{a}_{o,\mathrm{0}}^\dagger, 
  \delta\hat{a}_{o,\rm in} , \delta\hat{a}_{o,\rm in}^\dagger , 
		\delta\hat{a}_{t,\mathrm{vac}}, \delta\hat{a}_{t,\mathrm{vac}}^\dagger, \delta\hat{a}_{\mathrm{tm,vac}} , \delta\hat{a}_{\mathrm{tm,vac}}^\dagger
		)^\top,
	\end{aligned}
\end{equation}
as well as the matrices that encode the deterministic part of the QLE,
\begin{equation} \label{eq:M}
	M(t) = \begin{pmatrix} 
		- \frac{\kappa_e}{2} & 0 & 0 & -ig(t) & -ig^*(t) & 0 & 0 & 0 \\
		0 &  - \frac{\kappa_e}{2} & +i g^*(t) & 0 & 0& ig(t) & 0& 0 \\
		0& -ig(t) & i\delta_o-\frac{\kappa_o}{2} & 0 & 0 & 0 & 0 & 0 \\
		ig^*(t) & 0 & 0 & -i\delta_o-\frac{\kappa_o}{2} & 0 & 0 & 0 & 0 \\
		-ig(t) & 0 & 0 & 0 & i\delta_t - \frac{\kappa_{t}}{2} & 0 & -iJ & 0 \\
		0 & ig^*(t) & 0 &  0 & 0 & -i\delta_t -\frac{\kappa_t}{2} & 0 & iJ \\
		0 & 0 & 0 & 0 & -iJ & 0 & i\delta_{\mathrm{tm}} -\frac{\kappa_{\mathrm{tm}}}{2} & 0 \\
		0 & 0 & 0 & 0 & 0 & iJ & 0 & -i \delta_{\mathrm{tm}} -\frac{\kappa_{\mathrm{tm}}}{2}
	\end{pmatrix},
\end{equation}
and 
\begin{equation}\label{eq:K}
	K = 
	\begin{pmatrix}
		\sqrt{(1-\eta_e)\kappa_{e}} & \sqrt{\eta_e\kappa_{e}} & 0  & 0 & 0 & 0  \\
		0 & 0 & \sqrt{(1-\eta_o)\kappa_{o}} & \sqrt{\eta_o\kappa_{o}}  & 0 &0  \\
		0 & 0 & 0 & 0 & \sqrt{\kappa_{t}} & 0  \\
		0 & 0 & 0 & 0 & 0 &  \sqrt{\kappa_{\mathrm{tm}}} 
	\end{pmatrix}
	\otimes \mathbbm{1}_2,
\end{equation}
which keeps track on which modes the noise acts. 

\subsubsection{Input-Output-Theory}
In the experiment the pump field is turned on at $t = 0$ and kept on until $\tau_{\rm pulse}$. 
For the optical pump pulse with length $\tau_{\rm pulse} = $\SI{250}{ns} (\SI{600}{ns}, see main text), we reject a certain $\tau_{\rm{delay}}$ = \SI{50}{ns} (\SI{100}{ns}) from the beginning of pulse data. 
Since $\kappa_p \tau_{\rm delay} \gtrsim 1$ we may assume that after $\tau_{\rm delay}$ the system has approached its steady state and especially that the pump mode is in its steady state. Consequently we may assume that $g(t > \tau_{\rm delay}) \simeq g$ is constant over time. 
One important figure of merit is the multiphoton cooperativity $C = 4 g^2/\kappa_o\kappa_e$, a measure for coherent coupling versus the microwave and optical dissipation.
Efficient entanglement generation can be achieved with complete anti-Stokes scattering suppression, while below the parametric instability threshold, i.e. $C<1$.

The output fields of the CEO device are
\begin{equation}
\boldsymbol{f}_{\rm out}(t)
=
(
\hat{a}_{e,\rm out}(t), 
\hat{a}_{e,\rm out}^\dagger(t) ,
\hat{a}_{o,\rm out}(t) , 
\hat{a}_{o,\rm out}^\dagger(t)  
)^\top,
\end{equation}
which consist of a contribution which was entangled via the coherent interactions $\boldsymbol{v}$ and a contribution which has not interacted with the device $\boldsymbol{f}_{\rm in}$.
The output field $\boldsymbol{f}_{\rm out}$ will then propagate to the measurement device and is most economically represented within the framework of input-output theory~\cite{gardiner1985},
\begin{equation}
\label{eq:fout_fin}
\boldsymbol{f}_{\mathrm{out}}(t) = L \boldsymbol{f}_{\rm in}(t) - N \boldsymbol{v}(t),
\end{equation}
where we define the matrices
\begin{equation}\label{eq:N}
	N = (N_J , \mathbb{0}_4),
	\quad 
	\text{with}	
	\quad
	N_J =  
	\mathrm{Diag}(\sqrt{\eta_e\kappa_e},\sqrt{\eta_e\kappa_e},\sqrt{\eta_o\kappa_o},\sqrt{\eta_o\kappa_o}),
\end{equation}
and
\begin{equation} \label{eq:L}
	L 
	= 
	\begin{pmatrix}
		0  & 1 & 	0 & 	0  & 	0  & 	0  \\
		0  & 0  & 0  & 1 & 0  & 0 
	\end{pmatrix} 
	\otimes 
	\mathbb{1}_2.
\end{equation}
As all modes have reached steady state, the correlations in the output field may be obtained by going to Fourier domain.
Here we commit to following convention of the Fourier transformation
\begin{equation}\label{eq:fout}
    \hat{A}(\omega) 
    = 
    \frac{1}{\sqrt{2\pi}}\int_{-\infty}^{\infty}\mathrm{d}\omega \,e^{i\omega t} \hat{A}(t),
\end{equation}
with the hermitian conjugate
\begin{equation}
    (\hat{A}(\omega))^\dagger = A^\dagger(-\omega).
\end{equation}
Note that in this convention e.g. $[a_{e}(\omega), a_{e}^\dagger(\omega^\prime)] = \delta(\omega + \omega^\prime)$ are canonical pairs.

In our experiments, we focus on the correlations between the output propagating spectral modes of frequencies $\omega_e + \Delta\omega_e$ and $\omega_o  - \Delta\omega_o$ respectively for microwave and optical fields~\cite{Braunstein2005, Zippilli2015}.
We note that, due to energy conservation in the SPDC process, we only focus on microwave and optical photon pairs around resonances with anti-correlated frequencies, i.e. $\Delta\omega_e =\Delta\omega_o = \Delta\omega$.
For this reason,
we focus on the following vector of output fields in the rotating frame,
\begin{equation}
    \boldsymbol{f}_{\rm out}(\omega)
    =
    ( \hat{a}_{e,\rm out}(\omega), \hat{a}_{e,\rm out}^\dagger(-\omega) , \hat{a}_{o,\rm out}(-\omega) , \hat{a}_{o,\rm out}^\dagger(\omega) )^\top,
\end{equation}
 in the Fourier domain.
From Eq.~\eqref{eq:QLE_vec} we obtain
\begin{equation}\label{eq:QLE_sol}
	\boldsymbol{v}(\omega) = 
	\underbrace{\left[i\omega O - M\right]^{-1} \cdot K}_{=\mathcal{S}(\omega)} \cdot \boldsymbol{f}_{\rm in}(\omega),
\end{equation}
with
\begin{equation}
    O = 
    \mathrm{Diag}(1,-1,1,1) \otimes \sigma_z.
\end{equation}
Here we defined the vector of modes
\begin{equation} \label{eq:vo}
    \boldsymbol{v}(\omega)
    =
    (
    \hat{a}_e(\omega), \hat{a}_e^\dagger(-\omega),
    \hat{a}_o(-\omega), \hat{a}_o^\dagger(\omega),
    \hat{a}_t(\omega) , \hat{a}_t^\dagger(-\omega),
    \hat{a}_{\rm tm}(\omega), \hat{a}^\dagger_{\rm tm}(-\omega)
    )^\top,
\end{equation}
as well as the vector of input fields
\begin{equation}
\begin{split}
        \boldsymbol{f}_{\rm in}(\omega)
    & =
    (
  \delta\hat{a}_{e,\mathrm{0}}(\omega), \delta\hat{a}_{e,\mathrm{0}}^\dagger(-\omega),
  \delta\hat{a}_{e,\rm in}(\omega),
  \delta\hat{a}_{e,\rm in}^\dagger(-\omega), 
  \delta\hat{a}_{o,\mathrm{0}}(-\omega), 
  \delta\hat{a}_{o,\mathrm{0}}^\dagger(\omega), 
  \delta\hat{a}_{o,\rm in}(-\omega) , 
  \delta\hat{a}_{o,\rm in}^\dagger(\omega) , \\
  & \;\;\;\;\;\;
  \delta\hat{a}_{t,\mathrm{vac}}(\omega), 
  \delta\hat{a}_{t,\mathrm{vac}}^\dagger(-\omega), 
  \delta\hat{a}_{\mathrm{tm,vac}}(\omega) , 
  \delta\hat{a}_{\mathrm{tm,vac}}^\dagger(-\omega)
	)^\top
\end{split}
\end{equation}
in the Fourier domain. 

The output fields (see Eq.~\eqref{eq:fout_fin}) of the CEO device are straight forwardly obtained since in the Fourier domain Eq.~\eqref{eq:fout_fin} is algebraic,
\begin{equation} \label{eq:in_out_t}
	\boldsymbol{f}_{\rm out}(\omega) 
	=
	L \boldsymbol{f}_{\rm in}(\omega) +  N \boldsymbol{v}(\omega) 
	=
	(L + N \cdot [i \omega O - M]^{-1} \cdot K)\boldsymbol{f}_{\rm in}(\omega).
\end{equation}
The input noise operator correlations are given by,
\begin{equation}\label{eq:finfino}
	\langle \boldsymbol{f}_{\rm in}(\omega) \boldsymbol{f}^\dagger_{\rm in}(\omega^\prime) \rangle =  D \delta(\omega + \omega^\prime),
\end{equation}
with
\begin{equation}\label{eq:D}
	D = 
	\mathrm{Diag}(
	\underbrace{\bar{n}_{e,\mathrm{int}}+1,\bar{n}_{e,\mathrm{int}}}_{{\rm bath:} e},\underbrace{\bar{n}_{e,\mathrm{wg}}+1,\bar{n}_{e,\mathrm{wg}}}_{{\rm waveguide:} e},
	\underbrace{1,0}_{{\rm bath:}o},\underbrace{1,0}_{{\rm detector:}o},
	\underbrace{1,0}_{{\rm bath:} t},
	\underbrace{1,0}_{{\rm bath:} \mathrm{tm}}).
\end{equation}
We note that, in our experiments, the microwave waveguide remains in the ground state, with $\bar{n}_{e,\rm wg} = 0$.
The spectral correlations of different output field can be simply obtained analytically from
\begin{equation}\label{eq:Sfoutfout}
\langle \boldsymbol{f}_{\rm out}(\omega)\boldsymbol{f}_{\rm out}^\dagger(\omega^\prime) \rangle
=
\underbrace{\mathcal{S}(\omega)D \mathcal{S}^\dagger(-\omega)}_{\tilde{C}_{\boldsymbol{f}\boldsymbol{f}^\dagger}(\omega)} \delta(\omega+ \omega^\prime).
\end{equation}
Here we implicitly define the $4\times 4$ matrix of output mode correlations with a single entry reading
\begin{equation} \label{eq:SABdef}
    \langle \hat{a}_{\rm out}(\omega)\hat{b}_{\rm out}(\omega^\prime) \rangle = \tilde{C}_{ab}(\omega)\delta(\omega + \omega^\prime),
\end{equation}
where the operators $\hat{a}_{\rm out}(\omega),\hat{b}_{\rm out}(\omega)$ were chosen from components of $\boldsymbol{f}_{\rm out }(\omega)$ in Eq.~\eqref{eq:fout}.

\subsubsection{Covariance Matrix of Filtered Output Fields}
We will now consider a situation where we define output field modes from a windowed Fourier transformation.
Below we will then show that these are indeed the experimentally observed signals.
We start by defining the (dimensionless) hermitian output field quadrature pair~\cite{Zippilli2015},
\begin{eqnarray} \label{eq:def_Xn}
    \hat{X}_\alpha(\omega_n) & = & 
    \frac{1}{\sqrt{2T}}\int_{-T/2}^{T/2}\mathrm{d}\tau\,
    e^{i\omega_n \tau}
    \hat{a}_{\alpha,{\rm out}}(\tau) + \mathrm{h.c.}, \\
    \label{eq:def_Pn}
    \hat{P}_\alpha(\omega_n) & = & 
    \frac{1}{\sqrt{2T}i}\int_{-T/2}^{T/2}\mathrm{d}\tau\,
    e^{i\omega_n \tau}
    \hat{a}_{\alpha,{\rm out}}(\tau)  + \mathrm{h.c.},
\end{eqnarray}
which meets the canonical commutation relation
$[\hat{X}_{\alpha}(\omega_n),\hat{P}_\beta(\omega_m)] = i \delta_{nm}\delta_{\alpha\beta}$ where $\alpha = e,o$.
Due to the finite window of the Fourier transformation, the  frequencies $\omega_n = \frac{2\pi}{T}n$ becomes discrete.
The quadrature modes at discrete frequencies $\omega_n$ can now be rewritten in terms of the (dimensionful) output fields $\boldsymbol{f}_{\rm out}(\omega)$ from Eq.~\eqref{eq:in_out_t}, which are defined in the continuous Fourier domain.
Therefore the quadrature operators may be obtained by convolution with the a filter function $\mathcal{G}(\omega)$
\begin{eqnarray}
    \hat{X}_\alpha(\omega_n) & = & 
        \frac{1}{\sqrt{2}}
    \int_{-\infty}^\infty\mathrm{d}\omega\,
        \mathcal{G}(\omega_n - \omega)\hat{a}_{\alpha,{\rm out}}(\omega) 
    + \mathrm{h.c.} \\
    \hat{P}_\alpha(\omega_n) & = & 
    \frac{1}{\sqrt{2}i}
    \int_{-\infty}^\infty\mathrm{d}\omega\,
    \mathcal{G}(\omega_n - \omega)
    \hat{a}_{\alpha,{\rm out}}(\omega) 
    + \mathrm{h.c.}
\end{eqnarray}
Here the filter is
\begin{equation} 
	\mathcal{G}(\omega)
	=
	\frac{1}{\sqrt{2\pi}}\int_{-\infty}^\infty\mathrm{d}\tau\,e^{i\omega \tau} 
	\frac{\mathbb{1}_{[0,T]}(\tau)}{\sqrt{T}}
	=
	\sqrt{\frac{2}{\pi T}} \frac{\sin(\omega T/2)}{\omega},
\end{equation} 
which is obtained from a Fourier transformation of the unit function $\mathbb{1}_{[-T/2,T/2]}(t) = 1 (0)$ for $\vert t \vert \le T/2$ ($\vert t \vert > T/2$).
A bipartite Gaussian state is characterized by the $4\times4$ \textit{covariance matrix} (CM),
\begin{equation} \label{eq:QLE_cov}
    V_{AB}(\omega_n)
    =
    \frac{1}{2}\langle \{\delta \hat{A}(\omega_n), \delta \hat{B}(\omega_n) \}\rangle.
\end{equation}
Here we defined $\delta \hat{A} = \hat{A} - \langle \hat{A}\rangle$ an operator with zero mean $\langle \delta \hat{A}\rangle = 0$ and the quadratures from
\begin{equation}\label{eq:QLE_cov_ops}
\hat{A}(\omega_n),\hat{B}(\omega_n)\in \{ \hat{X}_e(\omega_n),\hat{P}_e(\omega_n),\hat{X}_o(-\omega_n),\hat{P}_o(-\omega_n) \}  
\end{equation}
and we also introduced the anti-commutator $\{\hat{A},\hat{B}\} = \hat{A}\hat{B} + \hat{B}\hat{A}$.
Note that the two-mode squeezing interaction results in correlation between frequency reversed pairs on the microwave $\omega_n$ and the optical side $-\omega_n$.
Since in our setting all first moments $\langle \hat{A}\rangle =0$ the evaluation of the covariance matrix in Eq.~\eqref{eq:QLE_cov} boils down to computing spectral correlations which are rewritten as
\begin{equation}
\begin{split}
\langle \hat{A}(\omega_n) \hat{B}(\omega_n) \rangle 
& =
\int_{-\infty}^\infty\mathrm{d}\omega\int_{-\infty}^\infty\mathrm{d}\omega^\prime\,
\mathcal{G}(\omega_n - \omega)
\mathcal{G}(\omega_n - \omega^\prime) \langle \hat{A}(\omega) \hat{B}(\omega^\prime) \rangle \\
& = 
\int_{-\infty}^\infty\mathrm{d}\omega\int_{-\infty}^\infty\mathrm{d}\omega^\prime\,
\mathcal{G}(\omega_n - \omega)
\mathcal{G}(-\omega_n - \omega^\prime) C_{AB}(\omega) \delta(\omega +\omega^\prime) \\
& =
\int_{-\infty}^\infty\mathrm{d}\omega\,  \mathcal{F}(\omega_n - \omega) C_{AB}(\omega),
\end{split}
\end{equation}
where we used the property $\mathcal{G}(-\omega) = \mathcal{G}(\omega)$ and defined the effective filter $\mathcal{F}(\omega) 
= 
\mathcal{G}(\omega)^2$.
Similar to Eq.~\eqref{eq:Sfoutfout}, we defined the quadrature correlations
\begin{equation} \label{eq:CABQLE}
C_{AB}(\omega) 
=
(C(\omega))_{AB}
= 
\frac{1}{2}
\left( U \tilde{C}_{\boldsymbol{f}\boldsymbol{f}^\dagger}(\omega)U^\dagger + (U \tilde{C}_{\boldsymbol{f}\boldsymbol{f}^\dagger}(\omega)U^\dagger)^\top \right)_{AB}.
\end{equation}
Here the unitary matrix $U = u \oplus u$, with 
\begin{equation} 
    u 
    =
    \frac{1}{\sqrt{2}}
    \begin{pmatrix}
        1 & 1 \\ -i & i
    \end{pmatrix},
\end{equation}
corresponds to a rotation of the mode operators into quadrature operators $(\hat{X}_\alpha, \hat{P}_\alpha)^\top =
u \cdot (\hat{a}_{\alpha,{\rm out}}, \hat{a}^\dagger_{\alpha,{\rm out}})^\top$.
The covariance matrix of the quadrature modes at the discrete frequencies $\omega_n$ is then obtained exactly by
\begin{equation} \label{eq:VAB_conv}
V_{AB}(\omega_n)
=
\int_{-\infty}^\infty\mathrm{d}\omega\,  \mathcal{F}(\omega_n - \omega) C_{AB}(\omega),
\end{equation}
where the quadrature correlations are convolved with an appropriate filter.

\subsection{Heterodyne Detection, Added Noise and Filtering}\label{subsec:output}

\subsubsection{Heterodyne Measurement}

Here we discuss the quadrature extractions from the equivalent linear measurement, e.g. balanced heterodyne detection, with excess added noise~\cite{dasilva2010}.
In the heterodyne detection, the output field $\hat{a}_{\mathrm{out}} e^{-i\omega_{j} t}$ ($j\in {e,o}$)
is mixed with a strong coherent local oscillator field $\hat{a}_{\mathrm{LO}}(t) = \alpha_{\mathrm{LO}} e^{-i\omega_{\mathrm{LO}} t}$ at a 50:50 beam-splitter,
where the output field from the two ports are sent to a balanced photo-detector, which results in a photon current that is proportional to
\begin{equation} \label{Iout_t}
	\hat{I}_{\rm out}(t)  	= 
	e^{-i\Delta_{\rm LO} t}\hat{a}_{{\rm out}}
 +
 \hat{a}_{{\rm out}}^\dagger e^{i\Delta_{\rm LO} t},
\end{equation}
in the limit of strong LO ($\alpha_{\rm LO}\gg 1$) with $\Delta_{\mathrm{LO}} = \omega_{\mathrm{LO}} - \omega_{j}$. 
We consider finite measurement interval of time $T$, on which we compute the windowed Fourier transformation of $\hat{I}_{\rm out}(t)$,
\begin{equation}\label{eq:Iout_o}
	\begin{split}
		\hat{I}_{\rm out}(\omega_n) 
		& = 
		\frac{1}{\sqrt{T}}\int_{0}^T\mathrm{d}\tau\,e^{i\omega_n \tau} \hat{I}_{\rm out}(\tau) 
		=
  \frac{1}{\sqrt{T}}\int_0^T\mathrm{d}\tau\,e^{i\omega_n \tau}
	(	
 e^{-i\Delta_{\rm LO}\tau} \hat{a}_{{\rm out}}(\tau)
 +
  e^{i\Delta_{\rm LO}\tau }\hat{a}_{{\rm out}}^\dagger(\tau)
    )
	\\
		& 
		=  
		a_{\rm out}(\omega_n-\Delta_{\rm LO})
  +
		a^\dagger_{\rm out}(\omega_n+\Delta_{\rm LO})  ,
	\end{split}
\end{equation}
where in a slight abuse of notation we define the dimensionless output fields $a_{\rm out}(\omega_n)$.
The reason why we are explicitly working the windowed Fourier transformation is, that despite being in a steady state during the measurement (see Sec.~\ref{subsec:output}), the Fourier transformed data has a rather broad bandwidth  ($\delta \omega_n  = \frac{2\pi}{T} \sim 5\mathrm{MHz}$ for a \SI{200}{ns} time window) due to the relatively short time of data collection $T = \tau_{\rm pulse} - \tau_{\rm delay} = 200 \,{\rm ns}$ (especially for \SI{250}{ns} optical pump pulse).
In the limit of long measurement times $T\rightarrow \infty$, the bandwidth will tend to zero and the following discussion as well as the results in Eq.~\eqref{eq:VAB_conv} will coincide with standard Input-Output treatment in the continuous Fourier domain.
In our experiments, we extract the quadratures of microwave and optical output field, by decomposing the heterodyne current spectra, in their real and imaginary parts which yields
\begin{equation}
    \hat{I}_{\rm out}(\omega_n)
    =
    \frac{1}{\sqrt{2}}
    ( 
    \underbrace{
        \hat{X}(\omega_n - \Delta_{\rm LO})
     +  
    \hat{X}(-\omega_n - \Delta_{\rm LO})
    }_{\hat{I}_{X,{\rm out}}(\omega_n)}
    +
    i
     \underbrace{ [
     \hat{P}(\omega_n - \Delta_{\rm LO})
     -
     \hat{P}(-\omega_n - \Delta_{\rm LO})
    ]}_{\hat{I}_{P,{\rm out}}(\omega_n)}
    ),
\end{equation}
where we define the quadrature output fields
$\hat{a}_{\rm out}(\omega_n) = (\hat{X}(\omega_n) + i \hat{P}(\omega_n))/\sqrt{2}$, in the same way as in Eq.~(\ref{eq:def_Xn}-\ref{eq:def_Pn}).

So far we have treated the photon current which result from a heterodyne measurement in terms of a time dependent hermitian operator $\hat{I}_{{\rm out}}(t)$.
In an actual experiment the heterodyne current is a real scalar $I(t)$ quantity which fluctuates in time and between different experimental runs. 
Taking taking the (fast) Fourier transform of this current and decomposing it in its real and imaginary parts then yields $I(\omega_n) = I_X(\omega_n) + i I_P(\omega_n)$.
The theory of continuous measurements and quantum trajectories \cite{walls2008,wiseman_milburn_2009} tells us how to connect the measured scalar currents with the current operators from input-output theory \cite{gardiner1985}
\begin{equation} \label{eq:IIcorr_class}
    \overline{I_A(\omega_n)I_B(\omega_m)} 
    =
    \frac{1}{2}
    \langle 
    \{ \hat{I}_{A,{\rm out}}(\omega_n),\hat{I}_{B,{\rm out}}(\omega_m)\}
    \rangle,
\end{equation}
where we define the statistical average  $\overline{\,\cdots\,}$ over many experimental runs.

\subsubsection{Realistic Measurements: Added Noise and Gain}
For the vacuum, the noise spectral density for both quadratures, are obtained by
\begin{equation} \label{eq:spectra_vac}
    S_{AA}(\omega_n) 
    =
    \langle \hat{A}(\omega_n) \hat{A}(\omega_n)\rangle_{\rm vac} = \frac{1}{2},
\end{equation}
for the hermitian operator $\hat{A} = \hat{X},\hat{P}$.
Note that due to the discreteness of the Fourier domain we do not have a Dirac delta as opposed to Eq.~\eqref{eq:SABdef}.
The noise spectrum of the heterodyne current is defined by $S_{II}(\omega) \equiv \overline{I(\omega_n)I(\omega_n)} = \langle \hat{I}_{\rm out}(\omega_n) \hat{I}_{\rm out}(\omega_n) \rangle$, where
\begin{equation}\label{eq:SII_bhd}
S_{II}(\omega_n)
= \frac{1}{2}\left(
S_{XX}\left(\omega_n-\Delta_{\mathrm{LO}}\right)+S_{PP}\left(\omega_n-\Delta_{\mathrm{LO}}\right)+S_{XX}\left(\omega_n+\Delta_{\mathrm{LO}}\right)+S_{PP}\left(\omega_n+\Delta_{\mathrm{LO}}\right)\right).
\end{equation}
Focusing on the part of the spectrum located around $\Delta_{\rm LO}$, 
\begin{equation}
S_{II}( \omega_n + \Delta_{\rm LO})
= \frac{1}{2}
\left(S_{XX}\left( \omega_n 
\right)+S_{{PP}}\left(\omega_n\right)+1\right),
\end{equation}
assuming $\Delta_{\rm LO}\gg \kappa_e,\kappa_o$.
This indicates the simultaneous quadratures measurements and added shot noise in the heterodyne measurements, even without experimental imperfections.
So far we have focused on the ideal theory of the measurement and disregarded additional unknown sources of noise as well as the connection to the actually measured quantities.
In practice, the \textit{decomposed measured quadratures contain additional uncorrelated excess noise}, e.g. due to the added noise in the amplification or due to propagation losses~\cite{caves1982}.
We model this by phenomenologically adding another uncorrelated noise process from an independent thermal reservoir and then multiplying by a gain factor which converts the number of measured photons to the actually monitored voltage.
To illustrate this we focus again on a single output port, and with the added noise current  $\hat{I}_{X/P,{\rm add}}(\omega_n)$  and the frequency dependent calibration gain $G_{\rm det}(\omega_n)$, where
\begin{eqnarray}
	\hat{I}_{X,{\rm det}}(\omega_n)=
	& 
	\sqrt{G_{\rm det}(\omega_n)}(\hat{I}_{X,{\rm add}}(\omega_n) + \hat{I}_{X,{\rm out}}(\omega_n)),\\
	\hat{I}_{P,{\rm det}}(\omega_n)=
	& 
	\sqrt{G_{\rm det}(\omega_n)}(\hat{I}_{P,{\rm add}}(\omega_n) + \hat{I}_{P,{\rm out}}(\omega_n)).
\end{eqnarray}
We thus obtain the detected heterodyne noise spectral density,
\begin{equation}\label{eq:SIItot}
\begin{split} 
S_{II,{\rm det}}(\omega_n + \Delta_{\rm LO}) 
= 
&
G_{\rm det}( \omega_n + \Delta_{\rm LO})
[S_{XX}\left(\omega_n \right)+S_{PP}\left(\omega_n\right)  \\
& 
+ 
\underbrace{ 
1
+
S_{I_XI_X,\mathrm{add}}( \omega_n+\Delta_{\rm LO})
+
S_{I_PI_P,\mathrm{add}}( \omega_n + \Delta_{\rm LO})
}_{=2N_{\rm add} }
],
\end{split} 
\end{equation}
where we define the spectra of the added noise $S_{I_{O} I_{O},\mathrm{add}}(\omega_n) = \langle \hat{I}_{O,{\rm add}}(\omega_n)
\hat{I}_{O,{\rm add}}(\omega_n)\rangle $.
\textit{The added noise $N_{\rm add}$ includes the excess vacuum noise from heterodyne measurement and the additional uncorrelated noise.}
Note that here the factor $\frac{1}{2}$ was absorbed in the detections gains.
The gain $G_{\mathrm{det}}(\omega_n)$ can be simply obtained on both microwave and optical side, from the cold measurements (optical pump off) with a known background.
We note that, Eq.~\eqref{eq:SIItot} 
lays the foundation of microwave and optical calibrations in our CEO device.

In our experiments, we place the LO on opposite sites around the mode resonances, i.e., 
\begin{equation}\label{eq:LOchoice}
    \Delta_{{\rm LO},e} = -\Omega_{\rm IF},
    \qquad
    \Delta_{{\rm LO},o} = \Omega_{\rm IF},
\end{equation}
where $\Omega_{\rm IF}>0$ is the intermediate frequency for down-mixing. 
The heterodyne output field can be obtained similar to Eq.~\eqref{eq:Iout_o},
\begin{equation}\label{eq:Iout}
\begin{aligned}
    \hat{I}_{\rm out,e}(\omega_n + \Omega_{\rm IF})
    &=\frac{1}{\sqrt{2}}
    [
    (
    \hat{X}_e(-\omega_n)
    +
    \hat{X}_e(\omega_n + 2 \Omega_{\rm IF}
    )
    +
    i (
     -\hat{P}_e(-\omega_n)
    +
    \hat{P}_e(\omega_n + 2\Omega_{\rm IF})
    ) 
    ],\\
    \hat{I}_{\rm out,o}(\omega_n + \Omega_{\rm IF})
    &=\frac{1}{\sqrt{2}}
    [
    \hat{X}_o(-\omega_n - 2 \Omega_{\rm IF})
    +
    \hat{X}_o(\omega_n)
    +
    i
    (
     -\hat{P}_o(-\omega_n-2 \Omega_{\rm IF})
    +
    \hat{P}_o(\omega_n)
    )
    ],
\end{aligned}
\end{equation}
with noise spectrum given by,
\begin{equation}
\begin{aligned}
S_{II,e}( \omega_n + \Omega_{\rm IF})
& =
\frac{1}{2}
(
S_{X_eX_e}\left( -\omega_n 
\right)+S_{{P_eP_e}}\left(-\omega_n\right))+  N_{e,\rm add},\\
S_{II,o}( \omega_n + \Omega_{\rm IF})
& =
\frac{1}{2}(
S_{X_oX_o}\left( \omega_n 
\right)+S_{{P_oP_o}}\left(\omega_n\right))
+ 
N_{o,\rm add}.
\end{aligned}
\end{equation}
We note that, Eq.~\eqref{eq:Iout} is adopted for \textit{field quadrature extraction (including the added noise) 
from the heterodyne measurement}, which reveals correlations in the quadrature histogram [cf. Fig.4 in the main text]. 
Despite of \textit{the reversed sign in the expected field quaduratures,  microwave and optical output photons appear at the same frequency in the noise spectrum, i.e. $\omega_n+\Omega_{\rm IF}$} [cf. Fig.2 d,e in the main text].

\subsubsection{Covariance Matrix from Realistic Heterodyne Measurements}
Here we briefly explain the procedure of the covariance matrix reconstruction from the heterodyne measurements.
\textit{The cross correlations of the detected heterodyne current spectra} can be obtained via,
\begin{equation}
    \mathcal{D}_{AB}(\omega_n)
    =
    \overline{\delta I_{A,{\rm det}}(\omega_n+\Omega_{{\rm IF}}) \delta I_{B,{\rm det}}(\omega_n+\Omega_{{\rm IF}})},
\end{equation}
where we define the centered current 
$\delta I_{O,\rm det} = I_{O,\rm det} - \overline{I_{O,\rm det}}$, with  
\begin{equation}
    I_{O,{\det}}(\omega_n) \in \{ I_{X_e,{\rm det}}(\omega_n), I_{P_e,{\rm det}}(\omega_n), I_{X_o,{\rm det}}(\omega_n), I_{P_o,{\rm det}}(\omega_n) \}. 
\end{equation}
Similar to Eq.~\eqref{eq:IIcorr_class}), 
we can obtain
\begin{equation}\label{eq:DAB}
\begin{aligned}
           \mathcal{D}_{AB}(\omega_n)
       & =
    \frac{1}{2}
    \langle 
    \{
    \delta \hat{I}_{A,{\rm det}}(\omega_n+\Omega_{{\rm IF}}),
    \delta \hat{I}_{B,{\rm det}}(\omega_n+\Omega_{{\rm IF}})
    \}
    \rangle\\
       & =
    \sqrt{G_{A,{\rm det}}( \omega_n+\Omega_{{\rm IF}}))G_{B,{\rm det}}( \omega_n+\Omega_{{\rm IF}}))}
    \biggl[
    \underbrace{
     \frac{1}{2}\langle\{ \delta \hat{A}(\omega_n),\delta \hat{B}(\omega_n) \}\rangle
    }_{=V_{AB}( \omega_n)}
    +
    N_{AB,{\rm add}}
    \biggr],
\end{aligned}
\end{equation}
where we define the diagonal added noise matrix $N_{AB,{\rm add}} = (N_{\mathrm{add}})_{AB} =  N_{A,{\rm add}}\delta_{AB}$ with the calibrated added noise $N_{\rm add}$ and detection gain $G_{A,{\rm det}}$.
\textit{
This equation establishes how the covariance matrix of the qudrature operators [cf. Eq.~\eqref{eq:VAB_conv}] is reconstruced from heterodyne measurements, and how they can be compared with the results from idealized standard input-output theory Eq.~\eqref{eq:QLE_cov}.}
For simplicity, in the main text we define the \textit{total measured covariance matrix including the added noise} as,
\begin{equation}
V_{AB,\rm meas}(\omega_n) 
= 
\mathcal{D}_{AB}( \omega_n )/\sqrt{G_{A,{\rm det}}( \omega_n +\Omega_{{\rm IF}}))G_{B,{\rm det}}( \omega_n +\Omega_{{\rm IF}}))},
\end{equation}
with $V_{AB,\rm meas}(\omega_n)=V_{AB}(\omega_n)+N_{AB,{\rm add}}$.

We note that,  in principle the location of both LOs can be arbitrary. 
As evident in Eq.~\ref{eq:DAB}, our choice of the LO configuration, i.e. $\Delta_{\rm LO,e}=- \Delta_{\rm LO,o} =- \Omega_{\rm IF}$, offers a simple solution to the quantification of the broadband quantum correlations, considering the limited detection bandwidth, frequency dependent gain, or microwave cavity frequency shift, which may result in the loss of quantum correlations during quadrature extractions in heterodyne measurements due to imperfect frequency matching.

\subsection{Entanglement Detection}

\subsubsection{Duan Criterion}
\label{subsubsection:Duan}
We will now discuss how show that the photons outgoing microwave and optical photons are indeed inseparable or entangled.
Our starting point is the covariance matrix which we defined in Eq.~\eqref{eq:QLE_cov} and measured as outline in Eq.~\eqref{eq:DAB}.
The experimentally measured covariance matrix is of the form
\begin{equation}
	V
	=
	\begin{pmatrix}
		V_e & V_{eo} \\ V_{eo} & V_o
	\end{pmatrix} 
	=
	\begin{pmatrix}
		V_{11} & 0 & \tilde{V}_{13} & \tilde{V}_{14} \\
		0 & V_{11} & \tilde{V}_{14} & -\tilde{V}_{13} \\
		\tilde{V}_{13} & \tilde{V}_{14} & V_{33} & 0 \\
		\tilde{V}_{14} & -\tilde{V}_{13} & 0 & V_{33}
	\end{pmatrix}.
\end{equation}
Since there is no single mode squeezing we have
$V_{22}= V_{11}$ and $V_{44}= V_{33}$.
For simplicity we have omitted the frequency argument $\omega_n$ of component. 
What we describe in the following will have to be repeated for every frequency component.
The off-diagonal part in the covariance matrix which encodes the two-mode squeezing can be written as
\begin{equation}
	V_{eo} 
\simeq 
V_{13}(\sin(\theta) \sigma_x + \cos(\theta) \sigma_z),
\end{equation}
where we define $ 
V_{13} = (\tilde{V}_{14}^2 + \tilde{V}_{13}^2)^{1/2}
$
and the mixing angle 
$\tan(\theta) = \tilde{V}_{14}/\tilde{V}_{13}$.
In our experimental setting $\tilde{V}_{14}$ maybe non zero e.g. due to small finite detunings $\delta_o$.
For the detection of inseparability, we employ the criterion introduced by Duan, Gidke, Cirac and Zoller~\cite{duan2000}.
This criterion states that if one can find local operations $U_{\rm loc} = U_e \otimes U_o$ such that the joint amplitude variance of $\hat{X}_{+} = (\hat{X}_{e} + \hat{X}_{o})/\sqrt{2}$ break the inequality,
\begin{equation} \label{eq:Delta_x+}
\Delta X^2_{+} = \langle U^\dagger_{\rm loc} \hat{X}^2_{+}U_{\rm loc}\rangle < 1/2,
\end{equation}
then the state is inseparable and, thus it must be concluded that it is entangled.

In this setting, it is enough to choose the local operations $U_{\rm loc} = U_{e} U_o$ to be a passive phase rotation on the optical mode only, with $U_e = \mathbb{1}$ and $U_o = e^{-i\varphi \hat{a}_o^\dagger \hat{a}_o}$, and phase rotation angle $\varphi$.
In the space of covariance matrices, this corresponds to the (symplectic) transformation $S_{\varphi} = \mathbb{1}_2 \oplus R_{\varphi}$, where we define the rotation matrix,
\begin{equation}
R_{\varphi}
	=
		\begin{pmatrix}
			\cos\left(\varphi\right) &  \sin\left(\varphi\right) \\
			-\sin\left(\varphi\right) & \cos\left(\varphi\right) 
	\end{pmatrix}.
\end{equation}
The local rotation of the phase $V(\varphi) = S_\varphi V S_\varphi^\top$ will act on the off diagonal part of the covariance matrix as,
\begin{equation}
 V_{ea}(\varphi)
= V_{13} (\cos(\theta - \varphi)\sigma_z 
+
\sin(\theta - \varphi)\sigma_x).
\end{equation}
With these local rotations the joint amplitude variance becomes
\begin{equation}\label{eq:Delta+Var}
	\Delta X^2_{+}(\varphi)
	= \langle(\hat{X}_e + \hat{X}_o\cos(\varphi) + \hat{P}_o \sin(\varphi))^2\rangle/2
	= V_{11} +V_{33} +2 V_{13} \cos(\theta-\varphi).
\end{equation}
We can similarly define the joint quadrature $\hat{P}_- = (\hat{P}_{e} - \hat{P}_{o})/\sqrt{2}$, where $\Delta P_-^2(\varphi) = 	\Delta X_+^2(\varphi)$.
The variance of the joint quadratures  $\Delta X_+^2(\varphi)$ and $\Delta P_-^2(\varphi)$ is minimized at the angle $\varphi_- = \theta - \pi$
\begin{equation}
	\Delta_{\rm EPR}^- = \Delta X_+^2(\varphi_-) + \Delta P_-^2(\varphi_-)= 2(V_{11}+V_{33} - 2V_{13}),
\end{equation}
which corresponds to the two-mode squeezing of microwave and optical output field, and the microwave-optics entanglement.
In addition, the joint quadrature variance is maximized at the angle $\varphi_+ = \theta$ and we obtain
\begin{equation}
	\Delta_{\rm EPR}^+ = \Delta X_+^2(\varphi_+) + \Delta P_-^2(\varphi_+) = 2(V_{11}+V_{33} + 2V_{13}),
\end{equation} 
which corresponds to the anti-squeezing.

\subsubsection{Logarithmic Negativity and Purity}
A mixed entangled state can be quantified by the logarithmic negativity,
\begin{equation}
E_{N}=\max \left[0,-\log \left(2 \zeta_{-}\right)\right],
\end{equation}
where $\zeta^-$ is the smaller symplectic eigenvalue of the partially time reverse covariance matrix and can be obtained analytically
\begin{equation}
    \zeta_-^2 
    =
    \frac{S - \sqrt{S^2 - 4 \mathrm{det}(V)}}{2}
\end{equation}
where we defined the Seralian invariant $S = \mathrm{det}(V_e) + \mathrm{det}V_o + 2 \mathrm{det}(V_{eo})$.
Furthermore the purity of a bipartite Gaussian state is given by
\begin{equation}
    \rho = \frac{1}{4\sqrt{\mathrm{det}(V)}},
\end{equation}
with $\rho=1$ for a pure state i. e. the vacuum state.


\section{Experimental Setup}\label{section:Setup}
\begin{figure}[t!]
	\centering
		\includegraphics[width=0.8\linewidth]{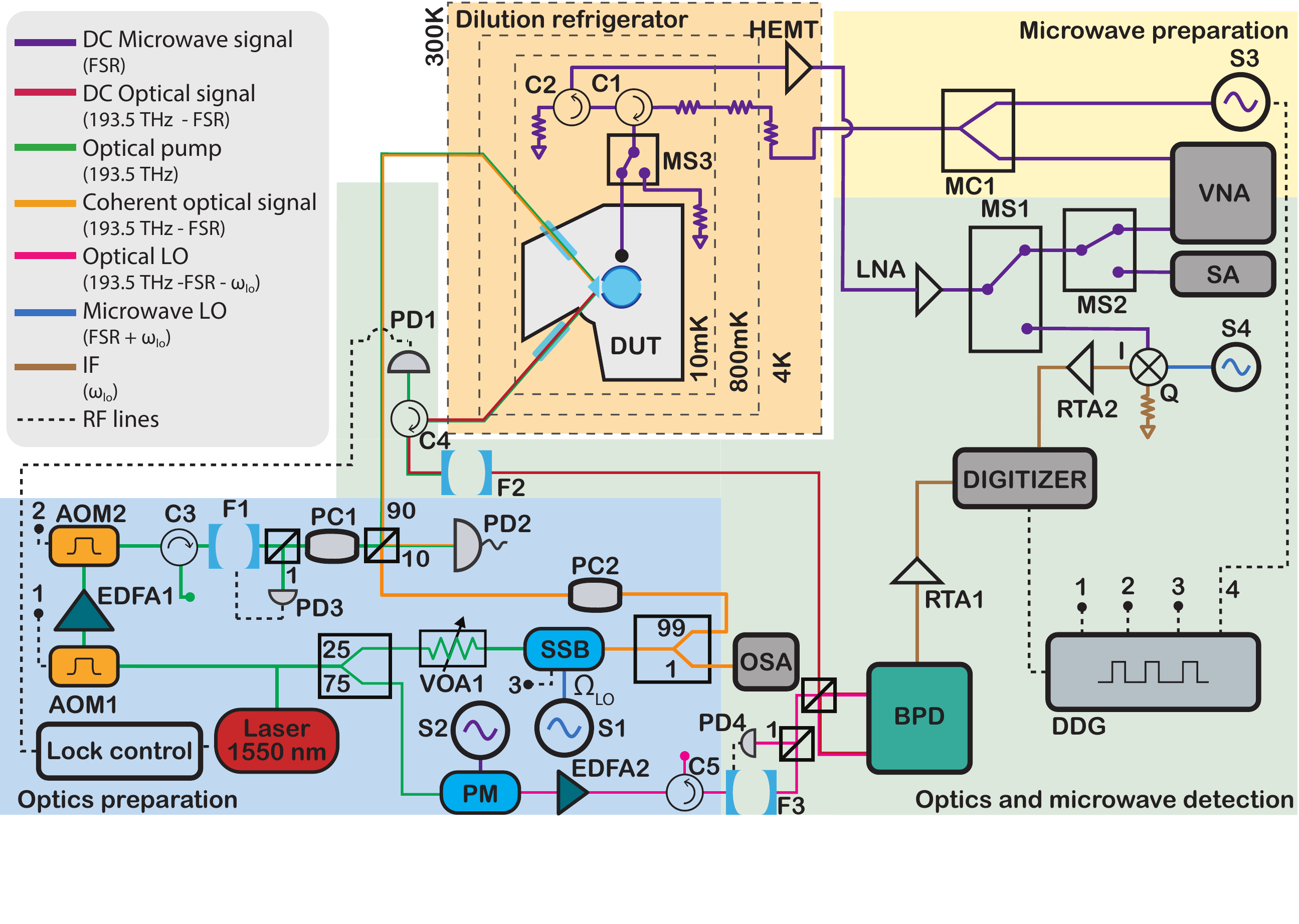}
 	\caption{\textbf{Experimental setup for two-mode squeezing measurements.} A tunable laser at frequency $\omega_p$ is initially divided equally in two parts, i.e. the optical pump and the optical signal together with the optical local oscillator (LO). Light from the optical pump path is pulsed via an acousto-optic modulator (AOM1) which produces \SI{}{ns}-pulses and shapes them for amplification via an Erbium-doped fiber amplifier (EDFA). The output from the EDFA is first filtered in time via AOM2 to remove the amplified spontaneous emission (ASE) noise and later in frequency via filter F1 ($\sim$\SI{50}{MHz} linewidth with \SI{15}{GHz} FSR) to remove any noise at the optical signal frequency (the reflected power is rejected by circulator C3). The filter F1 is locked to the transmitted power by taking 1\% of the filter transmission measured via photodiode PD3. The polarization of the final output is controlled via polarization controller PC1 before being mixed with the optical signal via a 90-10 beam splitter and sent to the dilution refrigerator (DR). The 10\% output from the beam splitter is monitored on a fast detector PD2 to measure the optical pump pulse power. The other half of the laser is again divided into two parts - 25\% for the optical signal and 75\% for the optical LO. The signal part is sent first to a variable optical attenuator VOA1 to control the power and then to a single sideband modulator SSB which produces the optical signal frequency at $ \omega_p - \text{FSR} $ and suppresses the tones at $ \omega_p $ and $ \omega_p + \text{FSR} $. 1\% of the optical signal is used to monitor the SSB suppression ratio via an optical spectrum analyzer OSA and 99\% is sent to the DR after being polarization controlled via PC2. The optical LO is produced via a phase modulator PM and detuned by $ \omega_\text{IF}/2\pi = \SI{40}{MHz} $. As the PM produces many sidebands, the undesired sidebands are suppressed via filter F3 ($\sim$\SI{50}{MHz} linewidth with \SI{15}{GHz} FSR), reflection is rejected by circulator C5. F3 is temperature-stabilized and locked to the transmitted power similar to F1. The optical LO is also amplified via EDFA2 before the optical balanced heterodyne. In the DR, the light is focused via a gradient-index (GRIN) lens on the surface of the prism and coupled to the optical whispering gallery mode resonator (WGMR) via evanescent coupling. Polarization controllers PC1 and PC2 are adjusted to efficiently couple to the TE modes of the optical WGMR. The output light is sent in a similar fashion to the collection grin lens. Outside the DR, the optical pump is filtered via filter F2 (similar to F3). The reflected light from F2 is redirected via C4 to be measured with PD1 which produces the lock signal for the laser to be locked to optical WGMR. The filtered signal is finally mixed with the optical LO and measured with a high speed balanced photo-diode BPD (\SI{400}{MHz}). The electrical signal from the BPD is amplified via RTA1 before getting digitized. On the microwave side, the signal is sent from the microwave source S3 which is connected to the DDG for accurately timed pulse generation (or from the VNA for microwave mode spectroscopy) to the fridge input line via the microwave combiner (MC1). The input line is attenuated with attenuators distributed between \SI{4}{K} and \SI{10}{mK} accumulating to \SI{50}{dB} in order to suppress room temperature microwave noise. Circulator C1 and C2 shield the reflected tone from the input signal and lead it to the amplified output line. The output line is amplified at \SI{4}{K} by a HEMT-amplifier and then at room temperature again with a low noise amplifier (LNA). The output line is connected to switch MS1 and MS2, to select between an ESA, a VNA or a digitizer measurement via manual downconversion using MW LO S4 (\SI{40}{MHz} detuned). Lastly, microwave switch MS3 allows to swap the device under test (DUT) for a temperature $T_{\textrm{\SI{50}{\ohm}}}$ controllable load, which serves as a broad band noise source in order to calibrate the microwave output line's total gain and added noise.}
	\label{fig:setup}
\end{figure}

The experimental setup is shown and described in SI Fig.~\ref{fig:setup}. The laser is split into three parts, including
an optical pulsed pump at frequency $\omega_p$, a continuous signal at $\omega_p - \rm FSR $ for the 4-port calibration  (cf. SI \ref{section:calibration}), and a continuous local oscillator (LO) at $\omega_p - \rm FSR + \Omega_{\rm IF}$ for the optical heterodyne detection. The optical signal and pump pulse are sent to the optical resonator of the electro-optical device (DUT) and the reflected light (with pump pulse rejected by a filter cavity) is combined on a 50:50 beam splitter with the optical local oscillator with subsequent balanced photodetection. 

Microwave input signals are attenuated at different temperature stages of the dilution refrigerator (4 K: 20 dB, 800 mK: 10 dB, 10 mK: 20 dB), and sent to the coupling port of the microwave cavity of the DUT. 
The reflected microwave signal is amplified and can then either be mixed with a microwave local oscillator of frequency $\rm FSR - \Omega_{\rm IF}$ and subsequently digitized, or directly measured by a vector network analyzer or a spectrum analyzer. 

We note that, \textit{the optical LO is on the right side of optical mode, while the microwave LO is on the left side of the microwave mode, with $\Omega_{\rm IF}/2\pi = 40 \rm MHz$.}
More details are in the caption of Supplementary Fig.~\ref{fig:setup}.






\section{Setup Characterization and calibration}\label{section:calibration} \label{sec:setup_characterization}
\begin{figure}[htbp]
	\centering
		\includegraphics[scale=0.5]{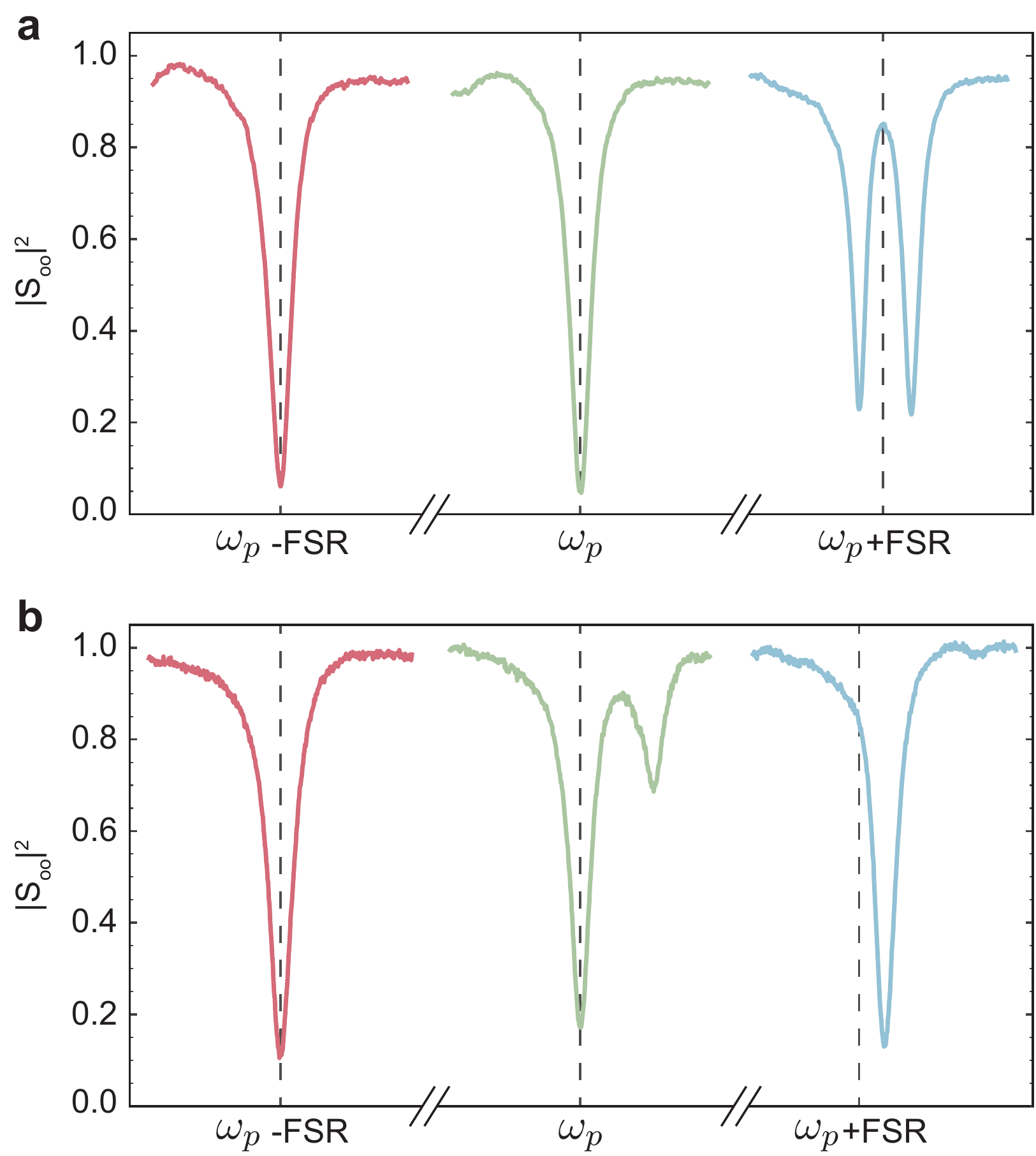}
 	\caption{\textbf{Optical mode spectra in reflection.} Normalized reflection intensity $|S_{oo}|^2$ spectra of optical modes $\hat{a}_o$, $\hat{a}_p$ and $\hat{a}_t$ in red, green and blue respectively. \textbf{a} (\textbf{b}) shows the optical mode spectra of the first (second) set of modes with the anti-Stokes and Stokes scattering ratio $\mathcal{S}=$ \SI{-10.3}{dB} (\SI{-3.1}{dB}). The dashed line marks the effective FSR between the pump mode $\hat{a}_p$ and the optical mode $\hat{a}_o$. The participation anti-Stokes optical mode $\hat{a}_t$ is suppressed for this effective FSR as marked by the dashed line over the blue mode.}
	\label{fig:OpticalModes}
\end{figure}

In the main manuscript, we show results from two different sets of optical modes shown in Fig.~\ref{fig:OpticalModes}. 
The main difference between these mode sets is the amount of suppression of the anti-Stokes scattering rate compared to Stokes scattering rate given by scattering ratio $\mathcal{S}$, which depends on
the mode hybridisation of the anti-Stokes mode~\cite{sahu2022,qiu2022}. 
The first set of optical mode (Fig.~\ref{fig:OpticalModes}a) from which we show most of our main results (main text Fig. 2, 3 and 5a) has $\mathcal{S}=$\SI{-10.3}{dB} on-resonance with an effective FSR = \SI{8.799}{GHz}. 
The last power sweep shown in main text Fig. 5b is measured with a second set of optical modes with a lower $\mathcal{S}=$\SI{-3.1}{dB} and a different effective FSR = \SI{8.791}{GHz} (Fig.~\ref{fig:OpticalModes}b). 
Despite it being the same optical resonator, the FSR for the second set of optical modes is slightly different, because of partial hybridisation of the optical pump mode which alters the working FSR between the optical pump and signal mode, see Fig.~\ref{fig:OpticalModes}.

In the following, we carefully calibrate the added noise due to the microwave detection chain at both these working FSRs (since microwave mode is parked at the working FSR). The added noise can be slightly different depending on frequency of measurement due to impedance mismatch and reflections between components in the microwave detection.

\subsection{Microwave added noise calibration} \label{subsection:MWaddednoise}
In the following, we carefully calibrate the slightly different added noise in the microwave detection chain at both frequencies.The impedance mismatch and reflections between components in the microwave detection chain can vary the added noise slightly as a function of frequency.
This added noise and corresponding gain due to a series of amplifiers and cable losses in the microwave detection chain is calibrated using a combination of a \SI{50}{\ohm} load, a thermometer and a resistive heater that are thermally connected. The microwave detection chain is identical for the signals from the \SI{50}{\ohm} load and the microwave cavity reflection, except for a small difference in cable length which we adjust for. 

To calibrate the detection chain, we heat the \SI{50}{\ohm} load with the resistive heater and record the amplified noise spectrum $P_{50\Omega}(\omega)$ as a function of temperature of \SI{50}{\ohm} load $T_{50\Omega}$. The output noise detected over a bandwidth $B$, $P_{50\Omega}$, as a function of $T_{50\Omega}$ is given as, 
\begin{equation}
    P_{50\Omega} = \hbar \omega_e G B \left[ \frac{1}{2} \coth\left(\frac{\hbar \omega_e}{2k_B T_{50\Omega}} \right) + N_{e,\text{add}}  \right],
    \label{eq:outputnoise_temp}
\end{equation}
with $\omega_e$ the center microwave frequency, $N_{e,\text{add}}$ ($G$) the added noise (gain) of the microwave detection chain, and $k_B$ the Boltzmann constant.

\begin{figure}[htbp]
	\centering
		\includegraphics[scale=0.5]{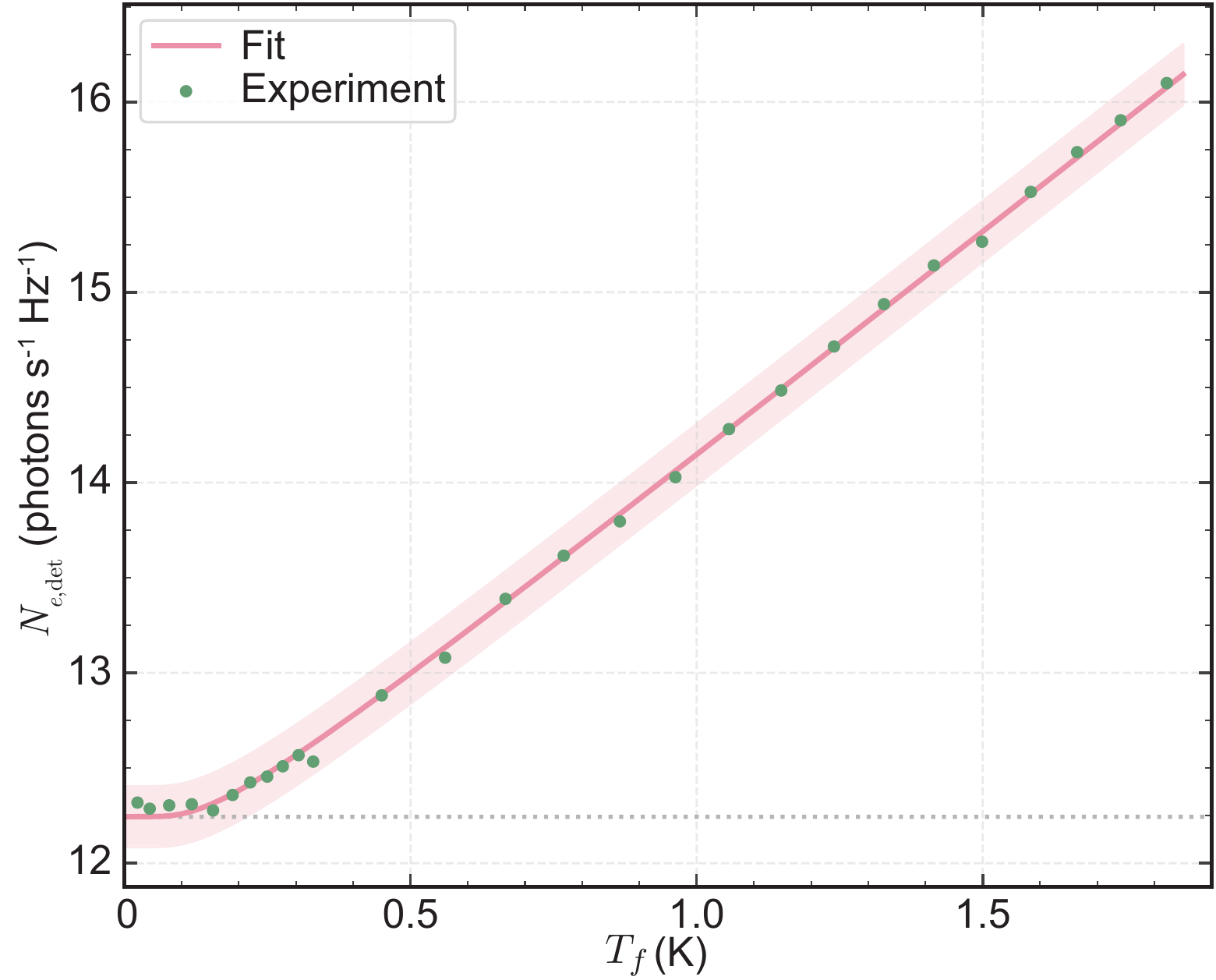}
 	\caption{\textbf{Characterization of the added noise in the microwave detection chain.} Measured output noise from a \SI{50}{\ohm} calibration load as a function of its temperature $T_f$. The measured noise is plotted in units of photons as $N_{det}^{50\Omega} = P_{50\Omega}/(\hbar \omega_e G B)$. 
 	The dashed line at the bottom represents the fitted vacuum noise level in addition to the added noise. 
 	The red line and shaded region represents the fit and the 95\% confidence interval around it.}
	\label{fig:HemtNoise}
\end{figure}

A bandwidth of \SI{11}{MHz} is selected around the region of interest to calculate $N_{e,\text{add}}$ and $G$.
For $\omega_e=$ \SI{8.799}{GHz}, we show the detected noise $N_{e,\text{det}} = P_{50\Omega}/(\hbar \omega_e G B)$ as a function of $T_{50\Omega}$ in SI Fig. \ref{fig:HemtNoise} along with a fit using Eq.~\ref{eq:outputnoise_temp}, with two fitting parameters $G$ and $N_{e,\text{add}}$. We note that, at $T_{50\Omega} = $ \SI{0}{K}, $N_{e,\text{det}} = N_{e,\text{add}} + 0.5$. 
Table~\ref{tab:mw_add_noise} (third row) shows the obtained added noise and gain for two frequencies of interest, i.e. $\omega_e/2\pi = $ \SI{8.799}{GHz} and $\omega_e/2\pi = $ \SI{8.791}{GHz}.

Next, we consider the difference in cable losses between the \SI{50}{\ohm} load and the microwave cavity, which are independently determined by measuring the microwave reflection from the microwave cavity and from the microwave switch directly before it. Including the cable losses, the effective added noise increases while the gain decreases for the reflected microwave detection, shown in Table~\ref{tab:mw_add_noise} (fourth row).

Finally, we consider an additional error due to the temperature sensor inaccuracy of $2.5\%$. Although this does not change the final $N_{e, \rm add}$ and $G$, it increases the uncertainty as shown in Table~\ref{tab:mw_add_noise} (fifth row). The error calculated in this section contributes to \textit{the systematic error} reported in the main text. 

\begin{table}[!htbp]
	\centering
	\caption{The added noise and gain in microwave detection chain ($1\sigma$ errors shown)}
	\begin{tabular}{|c|cc|cc|}
		\hline
		& \multicolumn{2}{c|}{8.799 GHz}                       & \multicolumn{2}{c|}{8.791 GHz}                       \\ \hline
		Detection Chain                                                                                & \multicolumn{1}{c|}{$N_{e,\rm add}$}              & G (dB)             & \multicolumn{1}{c|}{$N_{e,\rm add}$}              & G (dB)             \\ \hline
		\begin{tabular}[c]{@{}c@{}}50$\Omega$ load\\ (with fitting error)\end{tabular}                 & \multicolumn{1}{c|}{$11.74\pm0.08$} & $66.67\pm0.02$ & \multicolumn{1}{c|}{$11.76\pm0.09$} & $66.72\pm0.03$ \\ \hline
		\begin{tabular}[c]{@{}c@{}}MW cavity\\ (including cable loss)\end{tabular}                     & \multicolumn{1}{c|}{$13.09\pm0.09$} & $66.20\pm0.02$ & \multicolumn{1}{c|}{$13.16\pm0.10$} & $66.23\pm0.03$ \\ \hline
		\begin{tabular}[c]{@{}c@{}}MW cavity\\ (including temperature sensor uncertainty)\end{tabular} & \multicolumn{1}{c|}{$13.09\pm0.33$} & $66.20\pm0.12$ & \multicolumn{1}{c|}{$13.16\pm0.34$} & $66.23\pm0.12$ \\ \hline
	\end{tabular}
\label{tab:mw_add_noise}
\end{table}

\subsection{Optical added noise} \label{subsection:Optaddednoise}
Optical added noise is calculated via 4-port calibration of our device~\cite{sahu2022}. In this calibration, we measure the coherent response of our device through its 4 ports - optical input/output and microwave input/output. Sending an optical (or microwave) signal to the DUT in combination with a strong pump leads to stimulated parametric down-conversion (StPDC) process, which generates an amplified microwave (optical) coherent signal. We measure the 4 $S$-parameters of our device - microwave reflection ($S_{11}$), optics reflection ($S_{22}$), microwave to optics transmission ($S_{21}$) and optics to microwave transmission ($S_{12}$). The mean transduction efficiency between microwave and optics of the DUT is then calculated as,
\begin{equation}
    \eta = \sqrt{\frac{S_{12}S_{21}}{S_{11}S_{22}}}.
    \label{eq:transeff}
\end{equation}

We use the transduction efficiency and $N_{e,\text{add}}$ in the microwave detection chain from Sec.~\ref{subsection:MWaddednoise} to calculate the optical added noise. $N_{e,\text{add}}$ is firstly used to calculate the effective microwave detection gain (different from the one in Sec.~\ref{subsection:MWaddednoise}, because the microwave detection line used for the 4-port calibration uses analog downconversion and digitization, while the thermal calibration uses SA, see Fig. \ref{fig:setup}). 
The microwave gain, along with the (off-resonant) microwave reflection measurement, is used to calculate the microwave input loss. 
We can obtain the microwave signal power at the DUT, which allows us to calculate the output optical power of the DUT using the transduction efficiency. 
In conjunction with the measured output power at the end of the detection chain, the losses in the optical detection path and hence, the effective added noise with respect to the optical port of the DUT can be calculated. 
The calculated optical added noise is $N_{o,\text{add}} = 5.54\pm0.21(7.42\pm0.22)$ for $\omega_e = $ \SI{8.799}{GHz} (\SI{8.791}{GHz}).

\section{Data treatment}
In this section, we describe all the steps for the data treatment in detail, which includes the time domain analysis (Sec.~\ref{subsection:TDA}), the pulse post-selection due to setup drift (Sec.~\ref{subsection:PPS}), the frequency domain analysis (Sec.~\ref{subsection:FDA}), and the quadrature correlations (Sec.~\ref{subsection:QuadCor}).


\subsection{Time-domain analysis}\label{subsection:TDA}
\begin{figure}[htbp]
	\centering
		\includegraphics[scale=0.5]{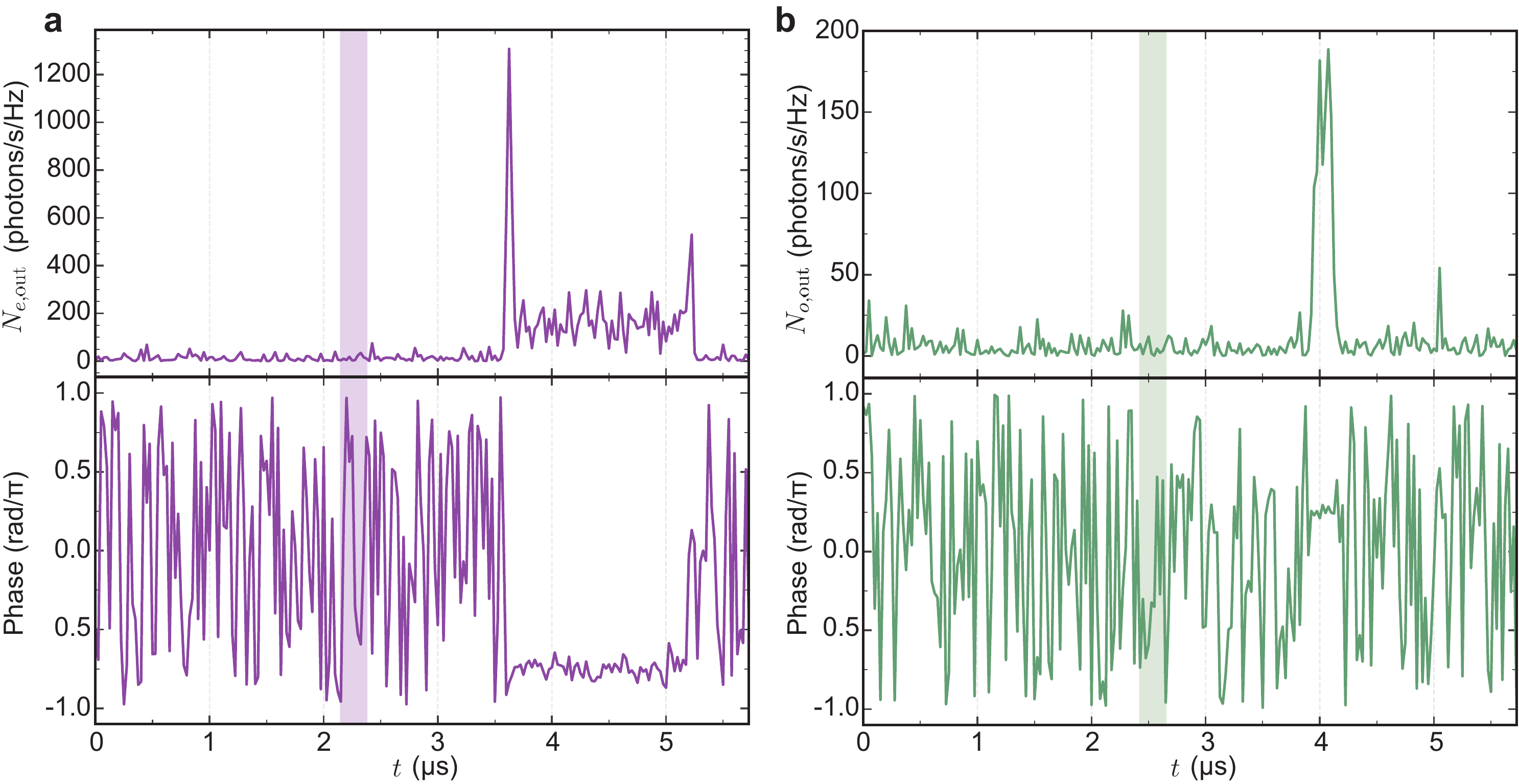}
 	\caption{\textbf{Downconverted output signal for a single measured pulse sequence.} \textbf{a (b)} show the measured microwave (optical) output signal downconverted at \SI{40}{MHz}. The shaded part in each case shows the region of the SPDC signal (the first optical pump pulse). For a single pulse, the SNR of a SPDC signal is too small to be seen. However, during the second optical pump pulse, a coherent response is seen in both signal outputs where the phase can be measured with high SNR for each single shot. }
	\label{fig:TimeDomainSI}
\end{figure}

Both microwave and optical signals are detected via heterodyne detection by mixing with a strong local oscillator that is $\sim$\SI{40}{MHz} detuned from respective mode resonance.
The output heterodyne signals are digitized using a digitizer at \num{1} GigaSamples/second. First, we digitally downconvert the digitized data at $ \omega_{\rm IF} = $ \SI{40}{MHz}. 
This yields the two quadratures $ I_{ X_{e/o},\rm det}(t)$ and $ I_{ P_{e/o},\rm det}(t)$ of the microwave or optical output signal record with \SI{40}{MHz} resolution bandwidth (using \SI{25}{ns} time resolution). Supplementary Fig. \ref{fig:TimeDomainSI} shows the calibrated output power ($I_{X_{e/o},\rm out}^2+I_{P_{e/o},\rm out}^2$)~[cf. Eq.~\ref{eq:Iout}] and the phase ($\arctan(I_{X_{e/o},\rm out}/I_{P_{e/o},\rm out})$) from a single pulse sequence. 
This includes the stochastic SPDC signals from a strong pump pulse, and the coherent StPDC signal from a weaker pump pulse together with a coherent microwave signal for calibration purposes. 
The SPDC signal produced by the first strong pulse is labeled by the shaded region for one single pulse, and the averaged output power over many pulses is shown in main text Fig. 2. 
The coherent microwave reflection and stimulated parametric downconverted optical signal are adopted to obtain the phases during the pulse.
We record this measured phase in both signal outputs during the second optical pump pulse for phase-drift correction in later post processing.

\begin{figure}[htbp]
	\centering
		\includegraphics[scale=0.5]{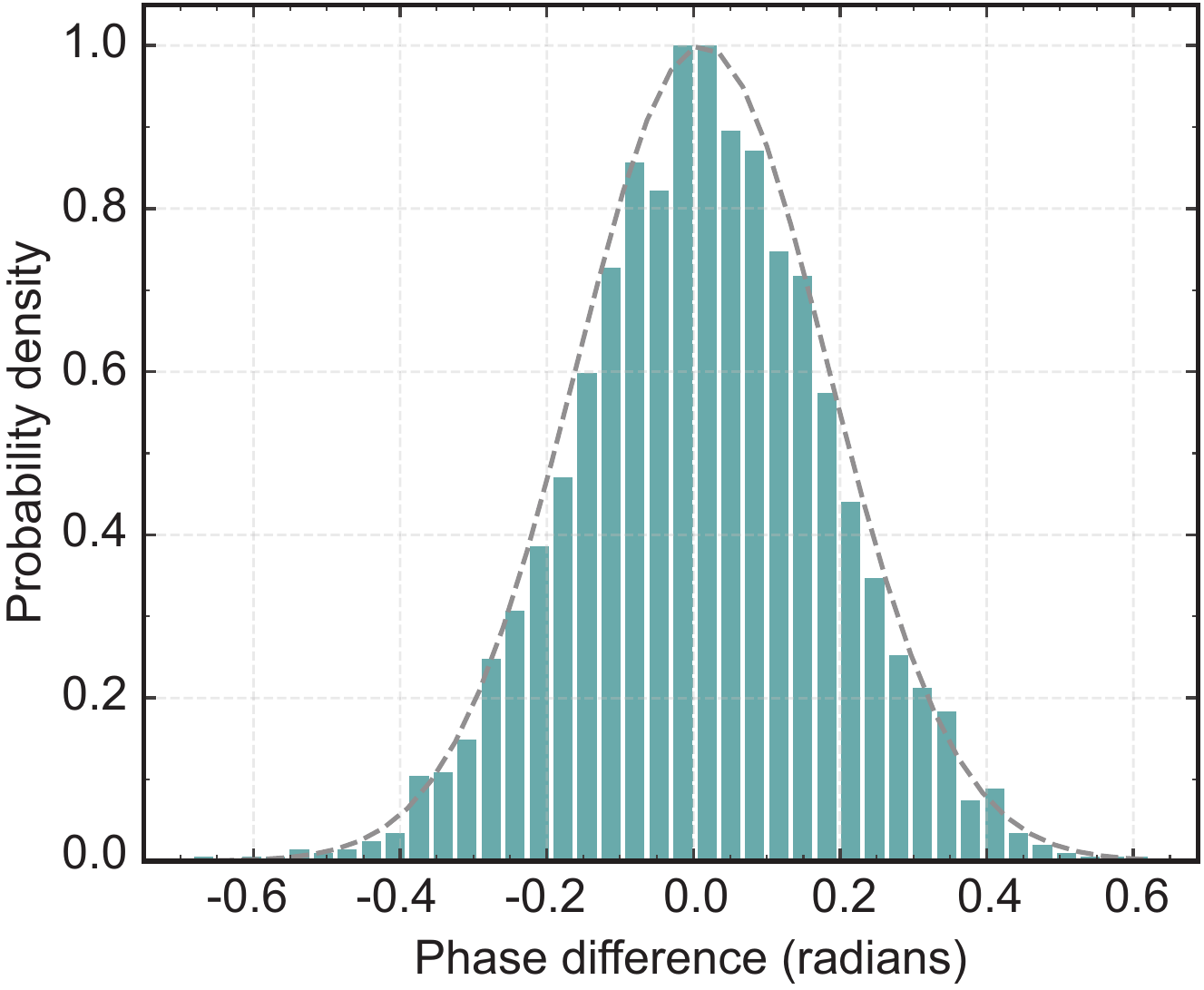}
 	\caption{\textbf{Accuracy of phase correction scheme.} The histogram shows the difference in the measured phase between the first and second optical pump pulse. Since we correct the phase in the first pump pulse based on the measured optical phase of the second optical pump pulse, the difference shows the limitations of this method. The grey dashed line is a normal distribution fit with variance \SI{0.17}{rad}.}
	\label{fig:PhaseCorr}
\end{figure}

In order to determine the accuracy of a phase correction for the first pump pulse based on the phase measurement during the second pump pulse, we send a continuous microwave signal during both pump pulses and recorded the phase of the converted optical pulse during the first and the second optical pump pulse. Supplementary Fig.~\ref{fig:PhaseCorr} shows the phase difference between the first and second optical pump pulse for 2500 trials along with a normal distribution fit. 
The fit variance for the distribution is \SI{0.17}{rad}. On a similar set of model data, applying a random phase variation of \SI{0.17}{rad} results in about 1.5-2.0\% loss of correlations [cf. Sec.~\ref{subsection:QuadCor}], whereas, we observe about 6-8\% loss of correlations in the experiments. 
The imperfection in phase correction does not completely explain the decreased quantum correlations, which might be due to other experimental instabilities, especially the optical pump laser lock.

\subsection{Pulse post-selection}\label{subsection:PPS}
In our experiments, we use three temperature-stabilized optical filters, which may drift slowly in time.
Two of them are used in the optical heterodyne detection.
In the signal path, one filter (F2 in Fig.\ref{fig:setup}) is used to filter the optical signal while reject the strong optical pump.
In the LO path, one filter (F3 in Fig.\ref{fig:setup}) is used to obtain  a clean optical LO tone, which is genearted by an electro-optic phase modulator (which produces multiple sidebands) and then amplified using an EDFA (which produces excess amplified spontaneous emissions). 

\begin{figure}[htbp]
	\centering
		\includegraphics[scale=0.5]{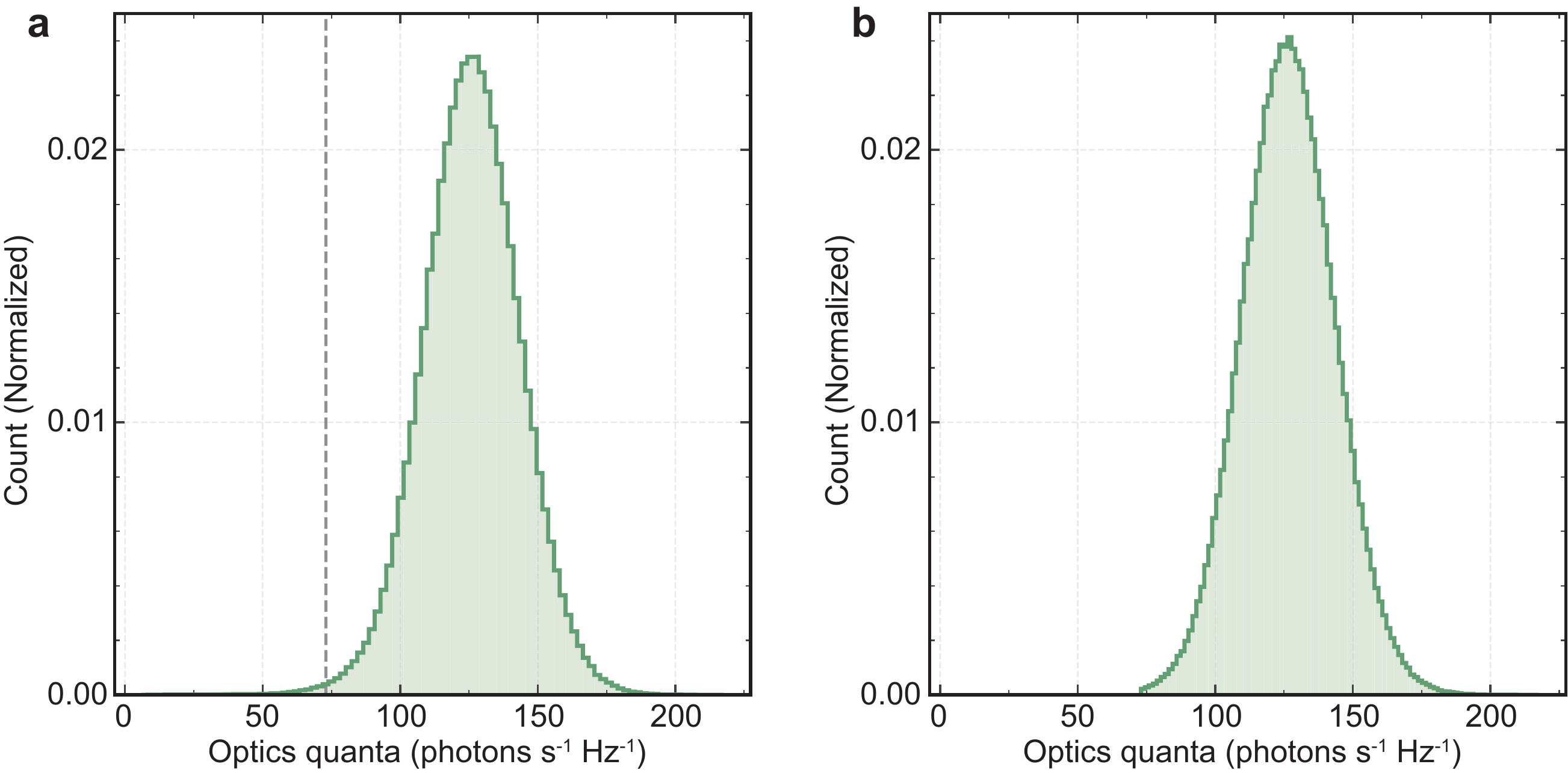}
 	\caption{\textbf{Histogram of the converted optical power.} The measured coherent optical power during the second optical pump pulse depends on the optical heterodyne gain and the received optical signal power. Both of these values can drift depending on the experimental setup's stability. \textbf{a} shows the normalized histogram of this measured optical power over all the collected pulses. The histogram has a tail on the lower end owing to the times when the heterodyne setup drifted. Filtering the points which do not meet a selected threshold (shown by the grey dashed line in \textbf{a}), we remove the instances where the setup had drifted and the optical heterodyne detection efficiency was compromised. The same histogram after removing such points is shown in \textbf{b}. }
	\label{fig:FilterHist}
\end{figure}

The slow filter drifts can be identified from the amplitude of the coherent optical signal produced via stimulated parametric downconversion during the second optical pump pulse, which drops due to either the decreased transmission after F2 or the reduced LO power after the F3. This is evident in the histogram of the converted optical power during the second optical pump pulse as shown in Fig. \ref{fig:FilterHist}a. The histogram is not symmetric and has a tail at the lower end.  
To filter out the instances of drifted heterodyne detection, we select a threshold (in this case marked by a dashed line in Fig \ref{fig:FilterHist}) and remove all pulses below the selected threshold along with 20 neighboring pulses (\SI{10}{s} in total time) before and after such instance. 
These numbers are chosen according to the filter drift and the filter temperature lock time-scales. 
After such filtering, usually about $10\%$ of the data is removed and the histogram of the converted optical power during the second optical pump pulse becomes symmetric as shown in Fig. \ref{fig:FilterHist}b. 


\subsection{Frequency domain analysis}\label{subsection:FDA}
As already mentioned in the main text, with the help of time-domain analysis, we select three different time-snippets to analyze the data in the frequency domain - before-pulse, on-pulse and post-pulse defined with respect to the first optical pump pulse. 
The main challenge in processing the data in frequency domain is the proper normalization of the measured output spectrum [cf. Eq.~\ref{eq:SIItot}]. 
The microwave reflection baseline is not flat because of slight impedance mismatches between different components in the microwave detection chain, with similar optical heterodyne shot noise floor due to the frequency dependent balanced detector gain. 
In addition, we observe slight shift of a few millivolts each time in the digitizer measurements when a new measurement is launched and the digitizer is reinitialized. 
Combined with the fact that the amplifier gain in the microwave detection chain as well as the optical heterodyne gain (due to optical LO power drift) may drift over a long time, an in-situ calibration of vacuum noise level is needed. 

In case of microwave, we need to first correct for the microwave reflection baseline distortion from impedance mismatch and then correct for the signal level shift caused by the digitizer. 
For the distorted baseline, we separately measure the microwave output spectrum when the microwave cavity is in its ground state (thermalized to \SI{7}{mK} at mixing chamber). This measurement is shown in Supplementary Fig \ref{fig:TimeDomainSI}a (gray) along with the measured before-pulse (cyan), on-pulse (purple) and after-pulse microwave noise spectrum (orange). 
Dividing the measured spectra with the cold cavity spectrum reveals a flat baseline Lorentzian noise spectra, however with an offset due to the digitizer drift.
To correct for this offset, we perform an in-situ vacuum noise calibration using the off-resonance (waveguide) noise in the before-pulse microwave noise spectrum.
An independent measurement of the microwave waveguide noise as a function of the average optical pump power (averaged over the full duty cycle) is shown in Supplementary Fig \ref{fig:Waveguidenoise}. 
The error bars ($2\sigma$ deviation) result from the microwave detection chain gain and the measurement instrument drift. 
The power law fit reveals that the microwave waveguide noise grows almost linearly with average optical pump power, and only deviates significantly from 0 for optical pump power $>$\SI{3}{\micro W}. 
As we work with average optical pump powers of $\ll$\SI{1}{\micro W}, we can safely assume the microwave waveguide noise to be zero. 
Therefore, we use the off-resonant waveguide noise for before-pulse microwave noise spectrum as an in-situ vacuum noise calibration. 

In case of optics, the optical detection is shot-noise limited, and the excess LO noise at the optical signal frequency is suppressed by more than \SI{40}{dB} using filter F1 in Fig.\ref{fig:setup}. 
We use the before-pulse optical noise spectrum as the vacuum noise level and normalize the optical on-pulse spectrum directly with the before-pulse in-situ calibration. Supplementary Fig \ref{fig:TimeDomainSI}b shows noise spectrum (without normalization) of the optical off-pulse (cyan), on-pulse (green), and the after-pulse (yellow). The signal during the optical pump pulse is clearly visible, and the noise level is identical before and after the optical pulse.

The normalized noise spectra for both microwave and optics are shown in main text Fig. 2d and 2e., where we can obtain the normalization gain [cf. Eq.~\ref{eq:SIItot}].

\begin{figure}[htbp]
	\centering
		\includegraphics[scale=0.5]{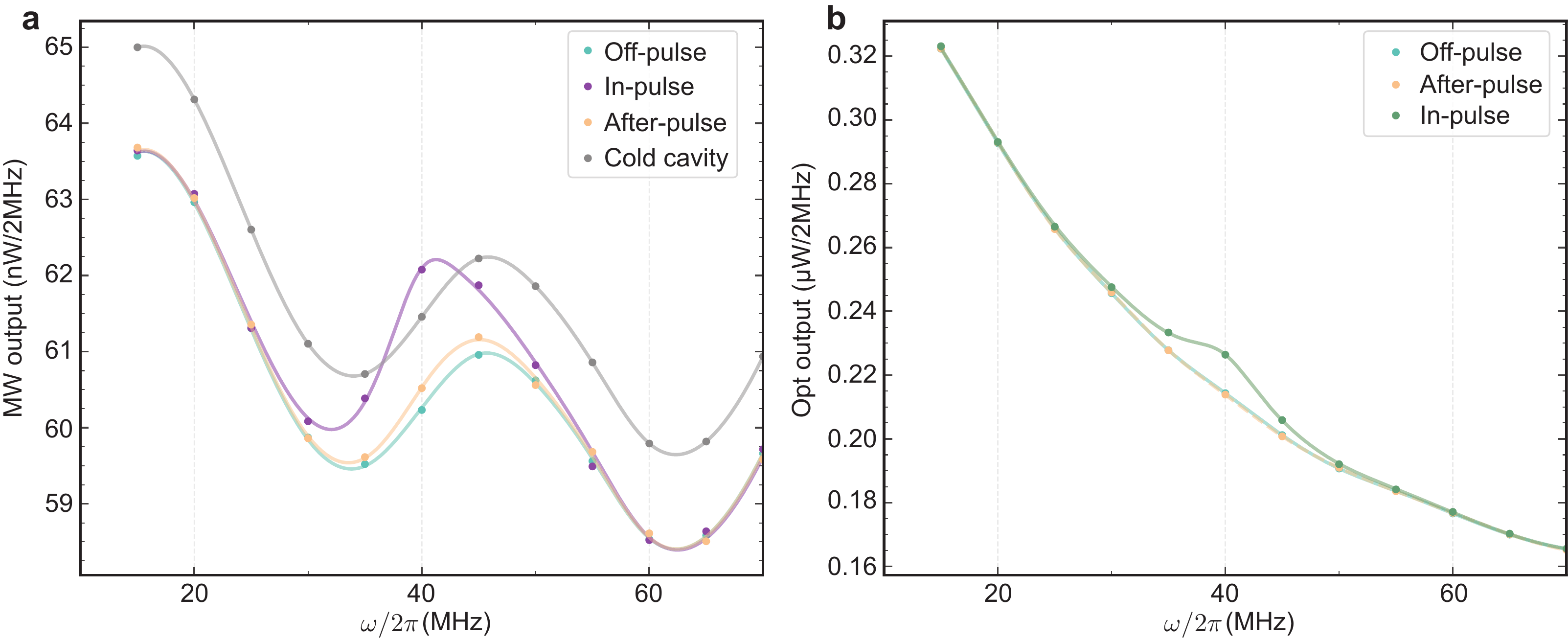}
 	\caption{\textbf{Spectra of output signals.} The microwave reflection baseline is not flat due to an impedance mismatch between different components in the microwave detection chain. As a result, the output power measured from amplified vacuum noise (from the cold microwave cavity) is not flat (shown in gray in \textbf{a}). Additionally, the digitizer in our setup has a different noise level each time it is started. As a result, the cold cavity baseline has an extra offset with respect to all other measurements. \textbf{a} also shows the measured output spectra for time region before (during, after) the first optical pulse shown in cyan (purple, orange). Similarly, \textbf{b} shows the output spectra for the optical output before (during, after) the first optical pulse in cyan (green, orange).}
	\label{fig:FreqDomainSI}
\end{figure}

\begin{figure}[htbp]
	\centering
		\includegraphics[scale=0.5]{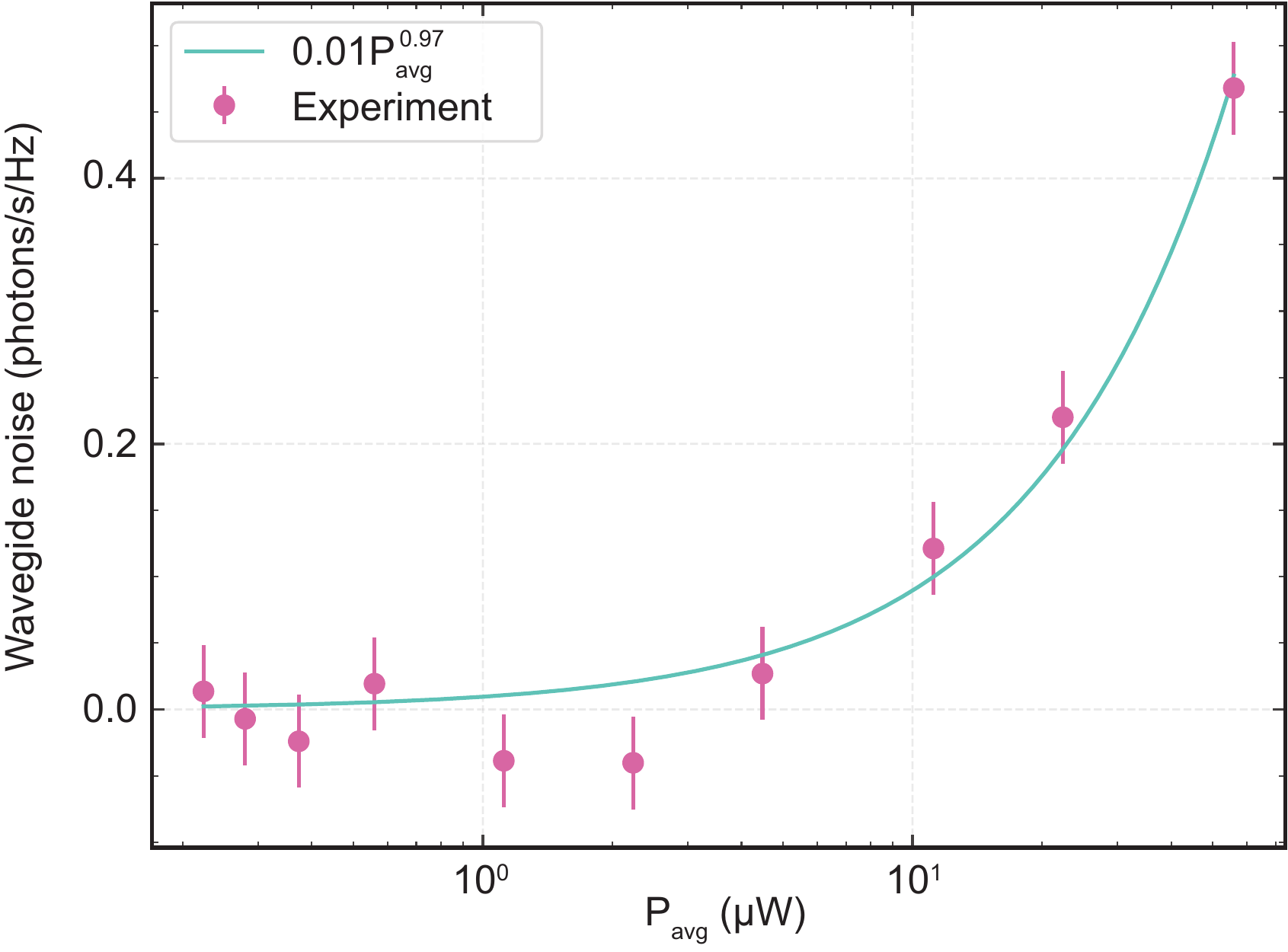}
 	\caption{\textbf{Microwave waveguide noise as a function of the average optical pump power.} The error bars represent $2\sigma$ error. The solid line is a power law fit. We find the power law is actually quite close to a linear function. }
	\label{fig:Waveguidenoise}
\end{figure}

\subsection{Joint-quadrature correlations}\label{subsection:QuadCor}
\textit{The detected output quadratures including excess added noise}, i.e. $\hat{I}_{X_e,\mathrm{out}}(\Delta \omega)$, $\hat{I}_{P_e,\mathrm{out}}(\Delta \omega)$,  $\hat{I}_{X_o,\mathrm{out}}(\Delta \omega)$, $\hat{I}_{P_o,\mathrm{out}}(\Delta \omega)$, can be obtained from the real and imagrinary parts in the discret Fourier transform of the photocurrent by normalizing to the detection gain [cf. Eq.~\ref{eq:Iout}].

Similar to Sec.~\ref{subsubsection:Duan}, we can define the joint  detected quadratures, by applying phase rotation on the optical ones,
\begin{equation}
	\begin{aligned}
	\hat{I}_{X,+} (\Delta \omega, \phi) &= \frac{\hat{I}_{X_e,\mathrm{out}}(\Delta \omega) 
		+ \left[\hat{I}_{X_o,\mathrm{out}}(\Delta \omega) \cos \phi - \hat{I}_{P_o,\mathrm{out}}(\Delta \omega) \sin \phi \right]}
	{\sqrt{2}},\\
	\hat{I}_{P,-} (\Delta \omega, \phi) &= \frac{\hat{I}_{P_e,\mathrm{out}}(\Delta \omega) - \left[\hat{I}_{X_o,\mathrm{out}}(\Delta \omega) \sin \phi + \hat{I}_{P_o,\mathrm{out}}(\Delta \omega) \cos \phi  \right]}
	{\sqrt{2}}.
	\end{aligned}
\end{equation}

To verify the non-classical correlation between the unitless quadrature variables for output microwave and optics field, i.e. $\hat{X}_e(\Delta \omega)$ \& $\hat{X}_o(-\Delta \omega)$ and $\hat{P}_e(\Delta \omega)$ \& $\hat{P}_o(-\Delta \omega)$, we can calculate the phase dependent joint quadrature variance [cf. Eq.~\ref{eq:Delta+Var}],
\begin{equation}
	\begin{aligned}
	\left<{\hat{X}^2_+} (\Delta \omega,\phi)\right> &= \left<
 {\hat{I}_{X,+}^2} (\Delta \omega,\phi)\right> - \frac{N_{e,\text{add}}+N_{o,\text{add}}}{2},\\
	 \left<{\hat{P}^2_-} (\Delta \omega,\phi)\right> &= \left<{\hat{I}_{P,-}^2} (\Delta \omega,\phi)\right> - \frac{N_{e,\text{add}}+N_{o,\text{add}}}{2}.
	\end{aligned}
\end{equation}

\begin{figure}[htbp]
	\centering
		\includegraphics[scale=0.5]{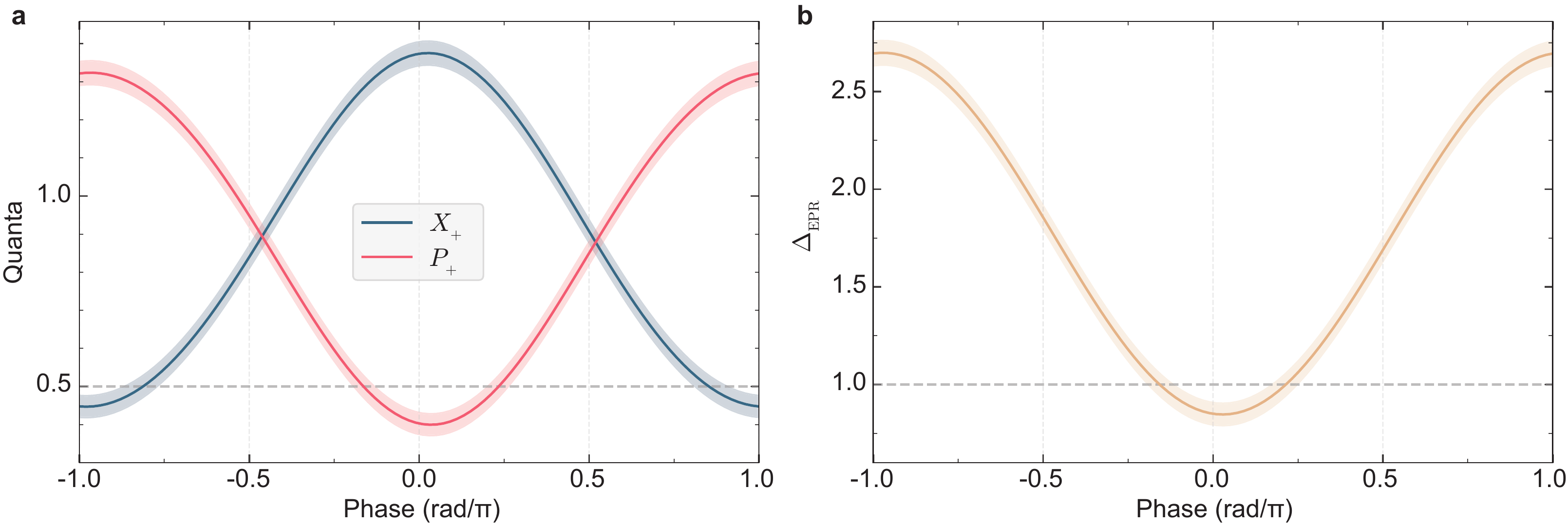}
 	\caption{\textbf{Joint quadrature correlations and $\Delta_\text{EPR}$.
  } \textbf{a.} 
  Joint quadratures at resonance $X_+ (\Delta \omega = 0)$ and $P_+ (\Delta \omega = 0)$ are plotted as a function of the local oscillator  phase $\phi$. \textbf{b.} $\Delta_\text{EPR}$ as a function of $\phi$. The shaded region in both plots represents the $2\sigma$ statistical error.}
	\label{fig:Duan_Phase}
\end{figure}

For $\Delta \omega = 0$, we plot the joint quadrature variance as a function of local oscillator phase in Fig. \ref{fig:Duan_Phase} (a). The shaded region represent the $2\sigma$ statistical error in the calculated joint quadrature variances. 
We note that, the statistical $1\sigma$ error of the variance for a Gaussian distributed data is given by $\sqrt{2}\sigma^2/\sqrt{N-1}$, where $N$ is the length of the dataset. 
The obtained resonant
$\Delta_\text{EPR}(0, \phi)$ is shown in Fig. \ref{fig:Duan_Phase}(b). The minimum and maximum of $\Delta_\text{EPR}(\phi)$ over the local oscillator phase are defined as $\text{min}[\Delta_\text{EPR}] = \Delta_\text{EPR}^-$ and $\text{max}[\Delta_\text{EPR}] = \Delta_\text{EPR}^+$. $\Delta_\text{EPR}^-<1$ indicates non-classical joint correlations and squeezing below vacuum levels. 

\begin{figure}[htbp]
	\centering
		\includegraphics[scale=0.5]{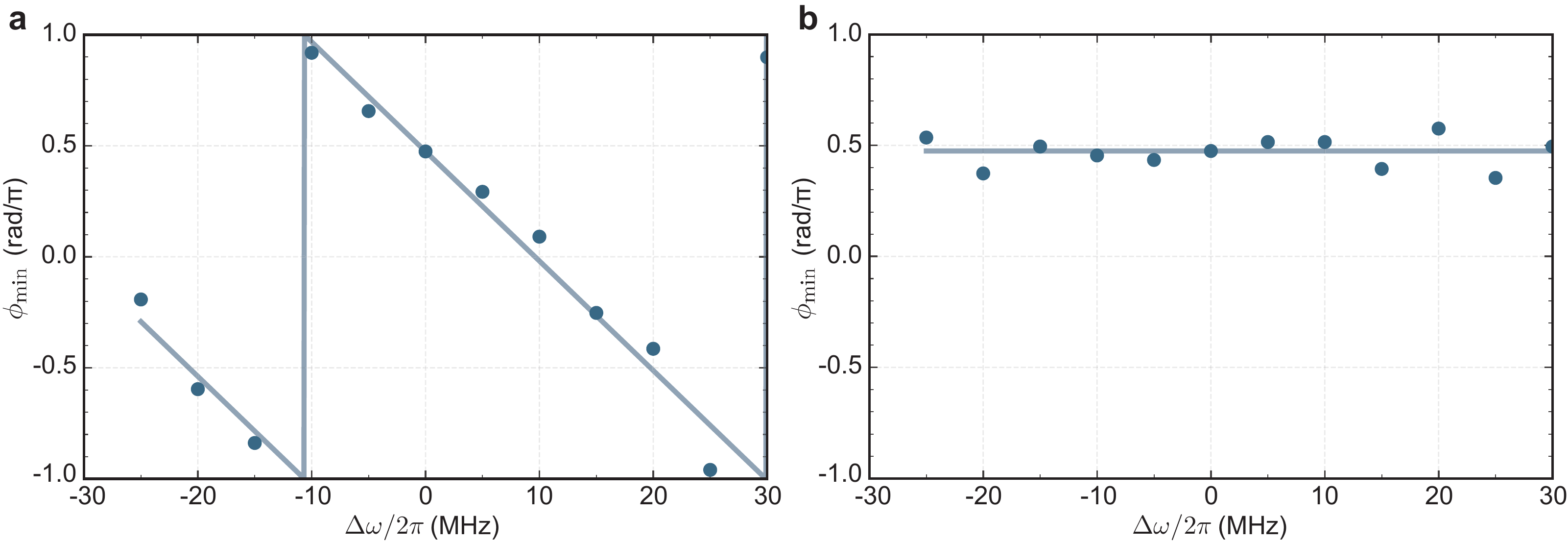}
 	\caption{\textbf{Effect of time delay between the microwave and optics signals.} The plots show the local oscillator phase $\phi_\text{min}$ which minimizes $\Delta_\text{EPR}(\Delta\omega, \phi)$ as a function of detuning frequency $\Delta \omega$. \textbf{a (b)} shows the case when the time difference of arrival between the microwave and optics signals was \SI{25}{ns} ($\approx$ \SI{0}{ns}). The solid lines are the linear fit to the experimental data.}
	\label{fig:Phase_vs_freq}
\end{figure}

The broadband phase that minimizes  $\Delta_\text{EPR}(\Delta\omega,\phi)$, i.e. $\phi_\text{min}(\Delta\omega)$, reveals the difference in arrival times (group delay) between the microwave and optical signal output (Supplementary Fig.~\ref{fig:Phase_vs_freq}a).  
After fixing the inferred time delay between the in-pulse arrival time of the microwave and optical signal, $\phi_\text{min}$ becomes independent of frequency detuning from the mode resonances. Thus, we adjust for the differences in arrival times by ensuring that the slope of $\phi_\text{min}$ with respect to detuning $\Delta \omega$ is minimized for all datasets we analyze, utilizing the broadband quantum correlations.

\section{Quadrature histogram raw data}
\begin{figure*}[h]
	\centering
	\includegraphics[scale=0.45]{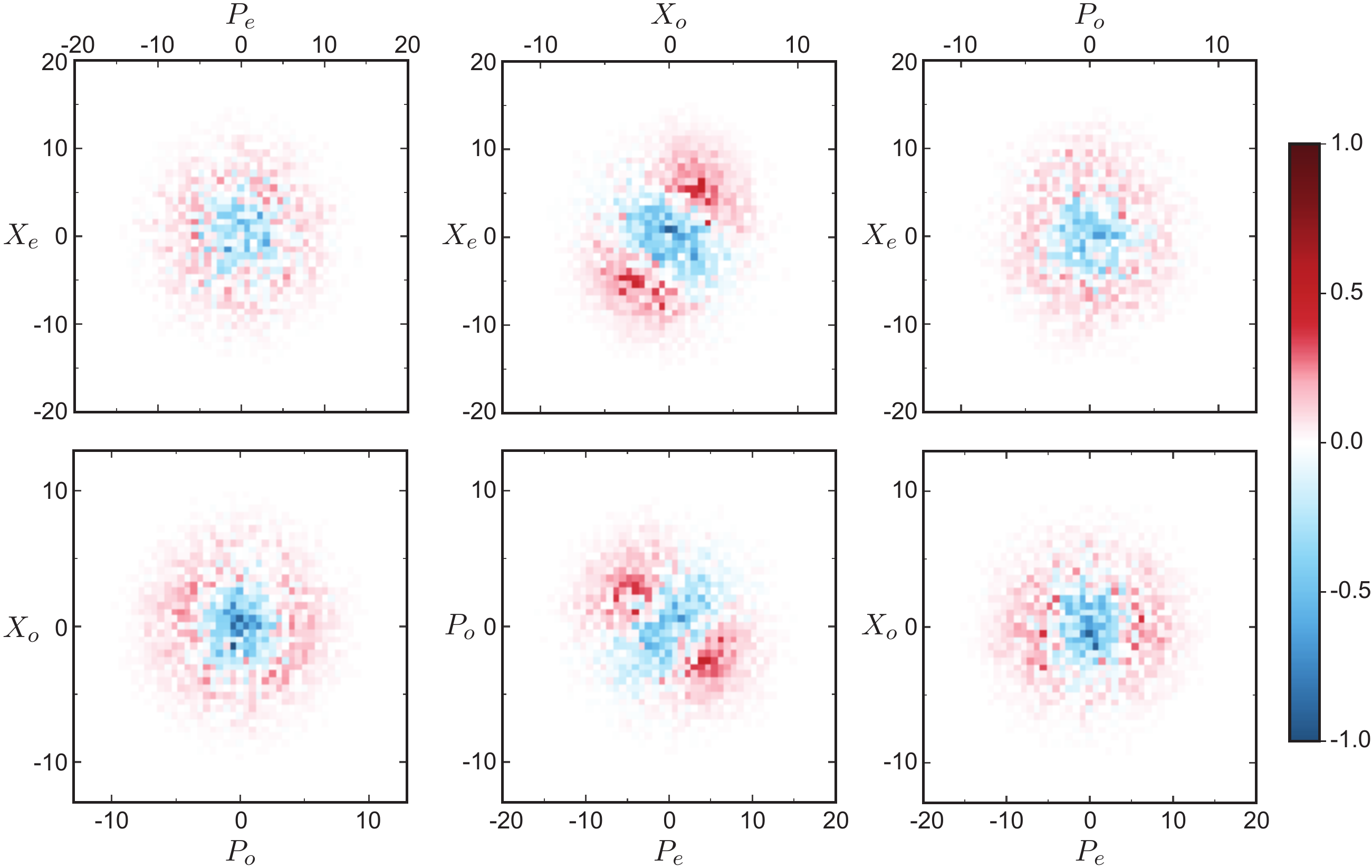}
	\caption{\textbf{Quadrature histogram raw data.} 
Normalized difference of the two-variable quadrature histograms obtained during and before the optical pump pulse based on the data shown in Figs.~2 and 3 of the main text.
	}
	\label{fig:afig1}
\end{figure*}

Fig.~\ref{fig:afig1} shows the normalized difference of the two-variable quadrature histograms obtained during and before the optical pump pulse based on the data shown in Figs.~2 and 3 of main text.
These unprocessed histograms directly show the phase insensitive amplification in each channel as well as the correlations in ($X_e$,$X_o$) and ($P_e$,$P_o$). Note however that - in contrast to the analysis in the main text - taking this difference does not lead to a valid phase space representation since also the vacuum noise of 0.5 together with the output noise of $0.026\pm0.011$ photons (due to the residual microwave bath occupancy right before the pulse) are subtracted, hence the negative values.

\section{Non-classical correlations with 600\,\lowercase{ns} long optical pump pulses}
\begin{figure*}[t]
	\centering
	\includegraphics[scale=0.5]{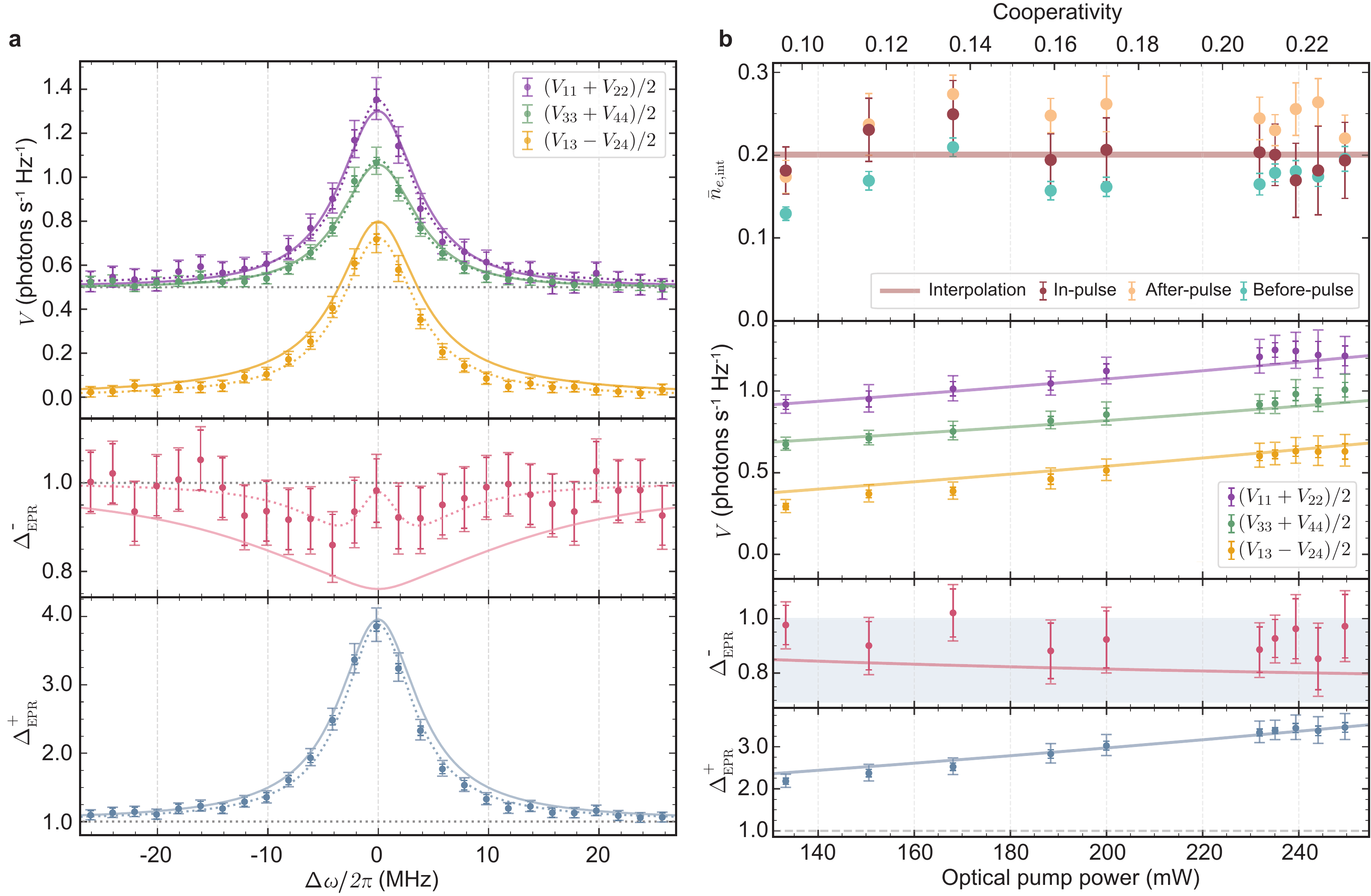}
	\caption{\textbf{Non-classical correlations vs.~optical pump power for 600\,ns long pump pulses.} 
\textbf{a}, Top panel, the average microwave output noise $\bar{V}_{11}$ (purple), the optical output noise $\bar{V}_{33}$ (green) and correlations $\bar{V}_{13}$ (yellow) as a function detuning based on 412500 measurements with a 2 Hz repetition rate. The solid lines represent the joint theory with fit parameters $C=0.22$ and in-pulse microwave thermal bath occupancy $\bar{n}_{e,\text{int}}=0.19\pm0.03$. The dashed lines are individual Lorentzian fits to serve as a guide to the eye. $\Delta_\text{EPR}^-$ ($\Delta_\text{EPR}^+$) in the middle (bottom) panel shown in red (blue) color are calculated from the top panel data and fits as described in the main text. The darker color error bars represent the $2\sigma$ statistical error and the outer (faint) error bars also include systematic errors. 
\textbf{b}, Power dependence of CM elements. The top panel shows the microwave mode thermal bath occupancy $\bar{n}_{e,\text{int}}$ for before-pulse, after-pulse and in-pulse regimes (marked in Fig.~2A) as a function of the peak optical pump power at the device and the corresponding cooperativity. The in-pulse $\bar{n}_{e,\text{int}}$ is obtained by the joint theory fit and approximated with a constant function (solid line). The middle panel shows the on-resonance mean CM elements based on the $\bar{n}_{e,\text{int}}$ from the top panel.
The bottom two panels show the \textit{on-resonance} squeezing $\Delta_\text{EPR}^-$ and anti-squeezing $\Delta_\text{EPR}^+$  calculated from the middle panel along with theory (solid lines). The darker color error bars represent the $2\sigma$ statistical error and the outer (faint) error bars also include systematic errors. All measured mean values are below the vacuum level and three power settings show a $>2\sigma$ significance for entanglement. }
	\label{fig:afig2}
\end{figure*}

Before experimenting with \SI{250}{ns} long optical pump pulses, we used \SI{600}{ns} long optical pump pulses. A sample measurement with a \SI{600}{ns} is shown in Fig.~\ref{fig:afig2}a similar to Fig.~3c of main text. Compared to \SI{250}{ns} long pulses, the main difference lies in the fact that $\Delta_\text{EPR}^-$ in the middle panel exhibits a double-dip shape because the correlations $\bar{V}_{13}$ have a wider bandwidth than the emitted noise spectra ($\bar{V}_{11}$ and $\bar{V}_{33}$), which are narrowed due to dynamical back-action~\cite{qiu2022}. Since in the measurement the correlations don't clearly overwhelm the emitted noise, interference between two Lorentzian functions of different widths (dashed line) leads to the specific shape of $\Delta_\text{EPR}^-$. Theory confirms this even though the shown theory curve (solid red line) does not exhibit the specific line-shape 
due to higher expected correlations compared to the experimentally observed values. 
These results indicate that $\bar{n}_{e,\text{int}}$ 
due to a \SI{600}{\nano s} optical pump pulse is large enough to prevent a clear observation of squeezing over the full bandwidth below the vacuum level ($\Delta_\text{EPR}^-<1$). As a result, we switched to \SI{250}{\nano s} optical pump pulses with higher statistics as shown in the main text. 

We also repeated the measurement with \SI{600}{ns} long pulses with different optical pump powers. Fig.~\ref{fig:afig2}b shows the measured pump power dependence with each data point based on 170000-412500 individual measurements each with a 2 Hz repetition rate. The microwave mode thermal bath occupancy $\bar{n}_{e,\text{int}}$ changes little as a function of the peak optical pump power at the device and is approximated with a constant function (solid maroon line in the top panel). The on-resonance mean CM elements scale with cooperativity and are in excellent agreement with theory (solid lines) based on the $\bar{n}_{e,\text{int}}$.
The \textit{on-resonance} squeezing $\Delta_\text{EPR}^-$ does not change significantly with cooperativity since both excess noise and correlations scale together with cooperativity. The anti-squeezing $\Delta_\text{EPR}^+$  scales up with cooperativity as expected. All but one measured mean values are below the vacuum level and three power settings show a $> 2 \sigma$ significance for entanglement. Note that this power sweep was conducted on a different set of optical modes with a different amount of anti-Stokes sideband suppression (see section \ref{sec:setup_characterization}).

\section{Error analysis}
The covariance matrix of the output field quadratures $V(\omega)$ can be directly calculated from the extracted microwave and optical quadratures from frequency domain analysis [cf. Eq.~\ref{eq:DAB}]
We simply rotate the optical quadratures with the phase that minimized the joint quadrature variance, and obtain the covariance matrix in the normal form.
We note that, the error in calculating the covariance matrix comes from two sources - the statistical error due to finite number of pulses, and the systematic error in the vacuum noise level calibration.
The detailed error analysis is described in the following subsections. We note that, \textit{the uncertainty in all the reported numbers in the main text corresponds to 2 standard deviation.}

\subsection{Statistical error}
\textit{The error in the calculation of bivariate variances} comes from the statistical uncertainties, arising from finite number of observations of a random sample. 
This error is the major component of our total error in diagonal covariance matrix elements. 
The error in calculating the variance of a sample distribution sampled from a Gaussian variable follows the Chi-squared distribution and is given as,
\begin{equation}
    \text{Var}(\sigma^2) = \frac{2\sigma^2}{N-1},
\end{equation}
where, $\sigma^2$ is the variance of sample distribution and $N$ is its size. 

In addition, \textit{the error in the covariance from a bivariate variable} is given by the Wishart distribution~\cite{wishart1928}. For a general bivariate covariance matrix $\Sigma$ given as,
\begin{equation}
    \Sigma = 
    \begin{pmatrix}
		\sigma_{11}^2 & \rho \sigma_{11}\sigma_{22} \\ \rho \sigma_{11}\sigma_{22} & \sigma_{22}^2
	\end{pmatrix}, 
\end{equation}
the variance of the covariance matrix is given by,
\begin{equation}
    \text{Var}(\Sigma) = \frac{1}{N-1}
    \begin{pmatrix}
		2\sigma_{11}^4 & (1+\rho^2) \sigma_{11}^2\sigma_{22}^2 \\ (1+\rho^2) \sigma_{11}^2\sigma_{22}^2 & 2\sigma_{22}^4
	\end{pmatrix}.
\end{equation}

\subsection{Systematic error}
Although, the systematic error in our measurements are not as significant, they still are a noticeable source of error. Here the error in calculating the covariance matrix results form the error in the estimation of the vacuum noise levels. More specifically, the error in determining the added noise due to the microwave and optical detection chain, as discussed in Sec.~\ref{section:calibration}. Propagating these systematic errors through the covariance matrix analysis is non-trivial, since calculating the error in variance of erroneous quantities is challenging. Therefore, we use a worst-case scenario approach to calculate the total error including the statistical error and the systematic error. We repeat the full analysis, including the statistical errors, for the lower and upper bound of the uncertainty range from the systematic errors for the microwave and optical added noise levels. Repeating the analysis expands the error bars in the calculated quantities. We take the extremum of all the error bars from all the repetitions of analysis to get the total error bar. 
We show both statistical error and the total error in the main text.


\let\oldaddcontentsline\addcontentsline
\renewcommand{\addcontentsline}[3]{}
\section*{References}
\let\addcontentsline\oldaddcontentsline

\renewcommand\bibname{References}
\addcontentsline{toc}{section}{\bibname}
\bibliographystyle{apsrev4-1}

\let\oldaddcontentsline\addcontentsline
\renewcommand{\addcontentsline}[3]{}
\bibliography{Final.bib}
\let\addcontentsline\oldaddcontentsline